\documentclass[10pt]{article}
\usepackage{amsmath}
\usepackage{graphicx}
\usepackage{color}
\usepackage{orcidlink}
\usepackage{hyperref}
\hypersetup{colorlinks=true, linkcolor=blue, citecolor=blue, urlcolor=blue}
\usepackage{caption}
\usepackage{subcaption}
\usepackage{setspace}
\usepackage{multirow}
\usepackage{multicol}
\usepackage{amssymb}
\usepackage{xcolor}
\RequirePackage[numbers,sort&compress]{natbib}
\begin{document}
\baselineskip=18pt

\newcommand{\rev}[1]{{\color{blue}{#1}}}
\newcommand{\revv}[1]{{\color{magenta}{#1}}}

\begin{center}
\setstretch{1.2}
\LARGE{\bf Observable Optical Signatures, Particle Dynamics and Epicyclic Frequencies of Mod(A)Max Black Holes }
\end{center}

\vspace{0.3cm}

\begin{center}
{\bf Faizuddin Ahmed\orcidlink{0000-0003-2196-9622}}\footnote{\bf faizuddinahmed15@gmail.com}\\
{\it Department of Physics, The Assam Royal Global University, Guwahati, 781035, Assam, India}\\
\vspace{0.2cm}
{\bf Ahmad Al-Badawi\orcidlink{0000-0002-3127-3453}}\footnote{\bf ahmadbadawi@ahu.edu.jo}\\
{\it Department of Physics, Al-Hussein Bin Talal University, 71111, Ma'an, Jordan}\\
\vspace{0.2cm}
{\bf Edilberto O. Silva\orcidlink{0000-0002-0297-5747}}\footnote{\bf edilberto.silva@ufma.br (Corresponding author)}\\
{\it Programa de P\'os-Gradua\c c\~ao em F\'{\i}sica \& Coordena\c c\~ao do Curso de F\'{\i}sica-Bacharelado, Universidade Federal do Maranh\~{a}o, 65085-580 S\~{a}o Lu\'{\i}s, Maranh\~{a}o, Brazil}
\end{center}

\vspace{0.2cm}

\begin{abstract}
\setstretch{1.5}
\noindent {\large In this work, we investigate the observable optical signatures of the Mod(A)Max black hole spacetime. We analyze key optical features, including the photon sphere, black hole shadow, and photon trajectories, and examine how these observables depend on the underlying geometric parameters, such as the electric charge and the Mod(A)Max coupling parameter. We further study the dynamics of neutral test particles in the vicinity of the black hole by deriving the effective potential within the Hamiltonian formalism. Using this potential, we obtain the specific energy and specific angular momentum for test particles on circular orbits of fixed radius, as well as the innermost stable circular orbit (ISCO), and explore how the geometric parameters influence these quantities and the ISCO radius. Finally, we derive the epicyclic (azimuthal, radial, and vertical) frequencies to analyze quasi-periodic oscillations (QPOs) exploring how the geometric parameters influences these and discuss their physical implications.}\\

\end{abstract}

\tableofcontents

\section{Introduction}\label{sec:1}

Black holes (BHs) are among the most fascinating predictions of Einstein's general theory of relativity (GR) \cite{AE1916,AE1935}, and their study remains a major focus in astrophysics. Disturbances in spacetime generated by extreme astrophysical events provide compelling evidence for the presence of BHs. The detection of gravitational waves by the LIGO and VIRGO collaborations in 2015 offered a spectacular confirmation of their existence \cite{GW2016,GW2017}. Further observational evidence has been provided by the Event Horizon Telescope (EHT), which achieved a historic milestone by capturing the first direct image of a BH at the center of the galaxy M87* \cite{EHTL1,EHTL4,EHTL5,EHTL6}. More recently, the EHT collaboration revealed the images of the supermassive BH Sagittarius A* (Sgr A*) at the center of the Milky Way \cite{EHTL12,EHTL14,EHTL15,EHTL16,EHTL17}. These breakthroughs in gravitational physics have significantly advanced our understanding of the cosmos and opened new avenues for exploring the nature of BHs.

The dynamics of test particles provide a powerful tool for probing black hole (BH) spacetimes, particularly in the strong-field regime near compact objects. Classical solutions, such as the Schwarzschild and Kerr BHs, have been thoroughly tested in both strong-field \cite{ss1,ss2,ss3,ss4} and weak-field \cite{GW2016,EHTL4} regimes. In recent years, Övgün and collaborators have made significant contributions to this area \cite{ss7,ss10}. Nonetheless, alternative and modified theories of gravity, including those involving scalar fields or a cosmological constant, remain open to astrophysical testing, with X-ray observations of compact objects providing important constraints \cite{ss11,ss12,ss13}. 

Circular orbits, in particular the innermost stable circular orbits (ISCOs), play a fundamental role in understanding the dynamics of test particles in the vicinity of black holes. 
Observations of accretion disks offer valuable constraints on black hole parameters and their physical properties \cite{ss14,ss15,ss16,ss17}. 
A large body of work has investigated particle dynamics around black holes in a variety of configurations within general relativity and modified theories of gravity. 
These studies demonstrate how the presence of external matter fields-such as dark matter, quintessence, and string clouds-can significantly influence the stability of marginally circular orbits 
\cite{FA1,FA2,FA7,FA9,FA11,FA12}.

Recent observational breakthroughs have strikingly confirmed the theoretical understanding of these orbits. The Event Horizon Telescope (EHT) images of M87* and Sgr A* provide direct visual evidence of the photon capture region, where light follows null geodesics that define the BH shadow. Interpreting these images relies on advanced general relativistic magnetohydrodynamic simulations that trace the polarized emission from hot plasma along these paths \cite{EHTL12}. These observations not only validate general relativity but also allow precision tests of the Kerr metric and constraints on alternative gravity theories \cite{ss20}. Additionally, long-term monitoring of stellar orbits around Sgr A* yields accurate BH mass measurements and offers opportunities to probe subtle relativistic effects \cite{ss21,ss22}. Collectively, the study of particle and photon motion provides a crucial link between spacetime geometry and empirical observations, establishing BHs as natural laboratories for testing fundamental physics.
 
Quasi-periodic oscillations (QPOs) observed in the X-ray spectra of BHs and neutron stars provide further probes of strong gravity. The epicyclic frequencies of neutral and charged test particles near compact objects are closely related to these oscillations \cite{ss44,ss45,ss46,ss47,ss48,ss49}. Observationally, QPOs manifest as low-frequency (LF) and high-frequency (HF) modes. LF QPOs are generally strong and stable, with slowly varying frequencies, while HF QPOs are weaker but encode crucial information about matter dynamics in strong gravitational fields \cite{ss50,ss51}. Some X-ray binaries exhibit both LF and HF QPOs simultaneously, indicating that they originate in distinct regions of the accretion disk. Moreover, Stuchlík and collaborators have studied the angular velocities of oscillating particles with respect to both proper time and static observers at infinity \cite{ss52,ss53}. Building on this work, Stuchlík \cite{ss54} analyzed three distinct QPO modes in the microquasar GRO 1655-40, comparing theoretical predictions with BH mass and spin values inferred from observations.This study examined several models, including relativistic precession, epicyclic resonance, tidal disruption, and warped-disk scenarios, providing important insights into the dynamics of accretion and oscillatory phenomena in BH spacetimes.. Several theoretical models have been proposed to explain the origin of quasi-periodic oscillations (QPOs), including (i) warped-disk, (ii) disk-seismic, (iii) resonance, and (iv) hot-spot models \cite{ss25}. Despite these efforts, the precise physical mechanism responsible for QPOs remains uncertain. Nonetheless, they remain a compelling area of research within general relativity and alternative theories of gravity. For a detailed discussion on the physical origins and observational characteristics of QPOs, see \cite{ss26}. Recent studies have further advanced our understanding of QPO phenomena around various black hole solutions \cite{ss27,ss28,ss29,ss30,ss31,ss32,ss32a,ss32b,ss32c,ss35,ss37,ss40,ss43,ss18,ss19,ss23}.

In this work, we investigate astrophysical observables associated with Mod(A)Max black holes, focusing on the photon sphere and black hole shadow. By analyzing the effective potential for null geodesics governing photon dynamics, we determine the photon sphere and shadow radii and examine how the black hole’s electric charge and the Mod(A)Max coupling parameter influence these quantities. Furthermore, we perform a comprehensive study of epicyclic frequencies in the spacetime of Mod(A)Max black holes, building on previous analyses of epicyclic motion \cite{KAB2018}. Our study provides a detailed characterization of particle dynamics around a Mod(A)Max black hole, highlighting the effects of mass, electric charge, and the Mod(A)Max coupling on orbital motion. Using the effective potential approach, we investigate the stability of circular equatorial orbits, derive exact analytical expressions for the specific energy and angular momentum of test particles, and explore how these parameters depend on black hole properties. We also analyze the innermost stable circular orbits (ISCOs), epicyclic oscillations, and periastron precession frequencies, offering key insights into particle motion in strong gravitational fields.

\section{Spherically Symmetric Mod(A)Max Black Hole: Revisiting Thermodynamics }\label{sec:2}

In a recent work~\cite{BEP2026}, the authors investigated new classes of topological black hole solutions in anti-de Sitter (AdS) spacetime within a novel nonlinear electrodynamics framework known as Modified Maxwell (ModMax) and its phantom counterpart, Modified anti-Maxwell (Mod(A)Max). In that study, the thermodynamic properties of these black holes were analyzed in detail, and significant physical aspects were discussed.  

In the present work, we focus on observable signatures of spherically symmetric Mod(A)Max black holes, including the photon sphere, black hole shadow, and the dynamics of neutral test particles in their vicinity. The line element describing a spherically symmetric Mod(A)Max black hole spacetime is given by~\cite{BEP2026}
\begin{equation}
    ds^2 = -f(r)\,dt^2 + \dfrac{dr^2}{f(r)} + r^2 (d\theta ^2 + \sin ^2{\theta }\,d\phi ^2),\label{metric}
\end{equation}
where the lapse function is given by\footnote[4]{One can generalize the black hole solution (\ref{metric}) for a dyonic Mod(A)Max black hole. In that case, the lapse function will be \(f(r)_{\rm dyonic\, Mod(A)Max}=1-\frac{2 M}{r}+\eta\,e^{-\gamma}\,\frac{(Q^2_e+Q^2_m)}{r^2}\), where $Q_e$ and $Q_m$, respectively are the electric and magnetic charges.}
\begin{align}
     f(r) = 1-\frac{2 M}{r}+\eta\,\frac{e^{-\gamma}\,Q^2}{r^2},\qquad A_{\mu} dx^{\mu}=-\frac{e^{-\gamma/2}\,Q}{r}\,dt.\label{function}
\end{align}
Here $M$ represents geometrical mass of the black hole, $Q$ is the electric charge, $\gamma$ is the ModMax's parameter, and the coupling constant $\eta=\pm\,1$ (plus sign corresponds to ModMax black hole \cite{DFA2021} and minus for phantom ModMax or Mod(A)Max black hole cases).

\begin{figure}[ht!]
\centering
\includegraphics[width=0.95\linewidth]{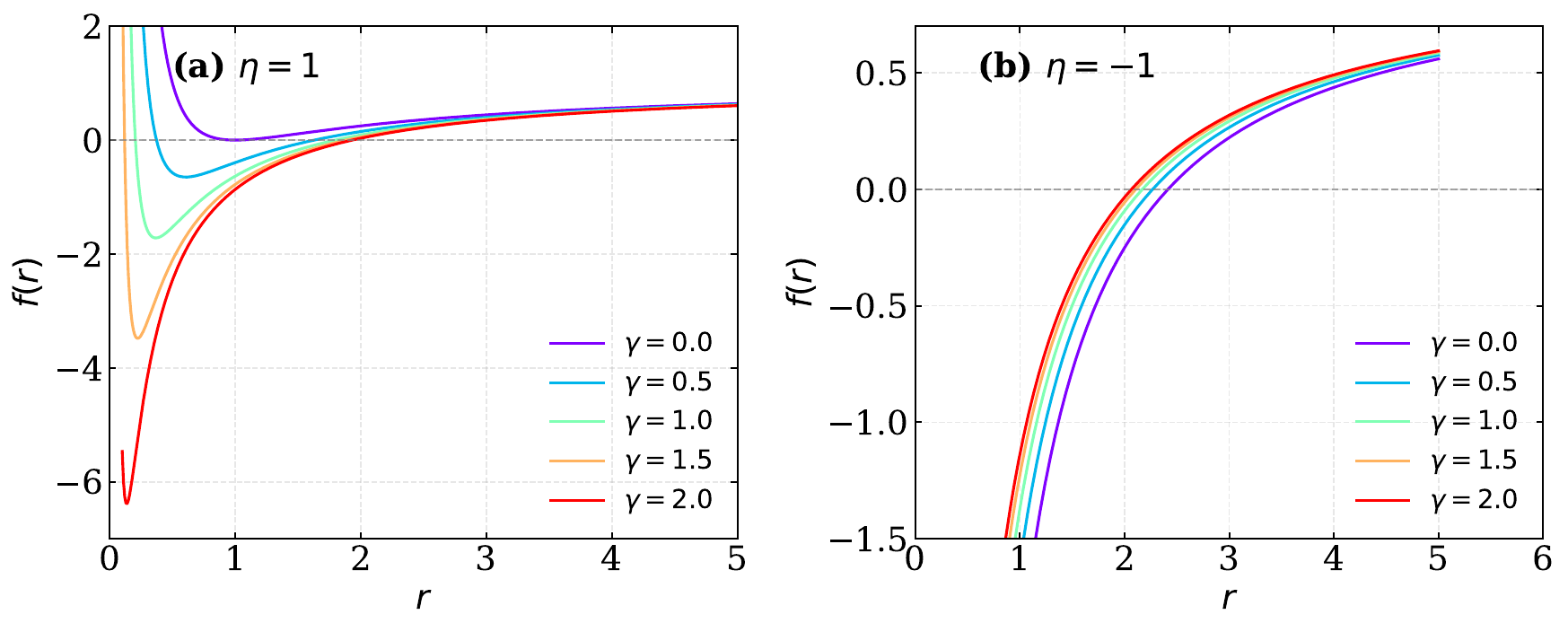}
\caption{Behavior of the metric function $f(r)$ for a Mod(A)Max black hole as a function of the radial coordinate $r$. Figure (a) illustrates the ModMax case ($\eta = +1$), while Fig. (b) shows the Mod(A)Max case ($\eta = -1$). Different curves correspond to different values of the ModMax parameter $\gamma$. The horizontal dashed gray line indicates $f(r) = 0$, where the event horizons are located. Here $M = 1$ and $Q = 1$.}
\label{fig:metric_function}
\end{figure}

In Fig.~\ref{fig:metric_function}, we display the radial profile of the metric function $f(r)$ for $\eta = \pm 1$. Figure \ref{fig:metric_function}(a) illustrates the ModMax case ($\eta = +1$): for fixed $Q = 1$, the solution can exhibit two horizons, one horizon (extremal case), or no horizons (naked singularity), depending on the value of $\gamma$. As $\gamma$ increases, the exponential suppression factor $e^{-\gamma}$ reduces the effective charge contribution, causing the metric function to approach the Schwarzschild limit and restoring the two-horizon structure.

Figure \ref{fig:metric_function}(b) shows the Mod(A)Max case ($\eta = -1$), where the phantom charge contribution leads to qualitatively different behavior. In this scenario, the solution describes a black hole with a single horizon for all values of $\gamma$ considered. The negative sign of $\eta$ reverses the effect of the charge term, effectively adding to the gravitational attraction rather than counteracting it. As a result, the metric function remains negative over a wider range of radii compared to the ModMax case, and the horizon radius increases with decreasing $\gamma$. These distinct horizon structures have important implications for the causal structure of the spacetime and the observable properties of Mod(A)Max black holes.

The metric function at radial infinity behaves as,
\begin{equation}
    \lim_{r \to \infty} f(r) \to 1 \label{asymptotic}
\end{equation}
which indicates that the lapse function is asymptotically flat. The horizons of the black hole are determined by the roots of the equation \(f(r) = 0\). These are given explicitly by
\begin{equation}
    r_{+} = r_h = M + \sqrt{M^2 - \eta\, e^{-\gamma} Q^2}, \qquad
    r_{-} = M - \sqrt{M^2 - \eta\, e^{-\gamma} Q^2}.
    \label{horizon}
\end{equation}
For the standard ModMax case (\(\eta = +1\)), the existence of an event horizon requires the constraint \(0 \leq e^{-\gamma} Q^2 \leq M^2\). On the other hand, for the phantom case (\(\eta = -1\)), there exists only a single horizon (see Fig~\ref{fig:metric_function}), and hence, no such restriction is necessary.

Therefore, the Hawking temperature at the horizon radius $r_h$ is given by
\begin{equation}
    T=\frac{f'(r_h)}{4\pi}=\frac{1}{4\pi\,r_h}\left(1-\eta\,e^{-\gamma}\,\frac{Q^2}{r^2_h}\right).\label{temperature}
\end{equation}
From the above expression, it is evident that the Hawking temperature of the ModMax black hole is lower than that of its phantom counterpart, namely the Mod(A)Max black hole. In other words, \(T_{\rm ModMax} < T_{\rm Mod(A)Max}.\) 

\begin{figure}[tbhp]
    \centering
    \includegraphics[width=0.45\linewidth]{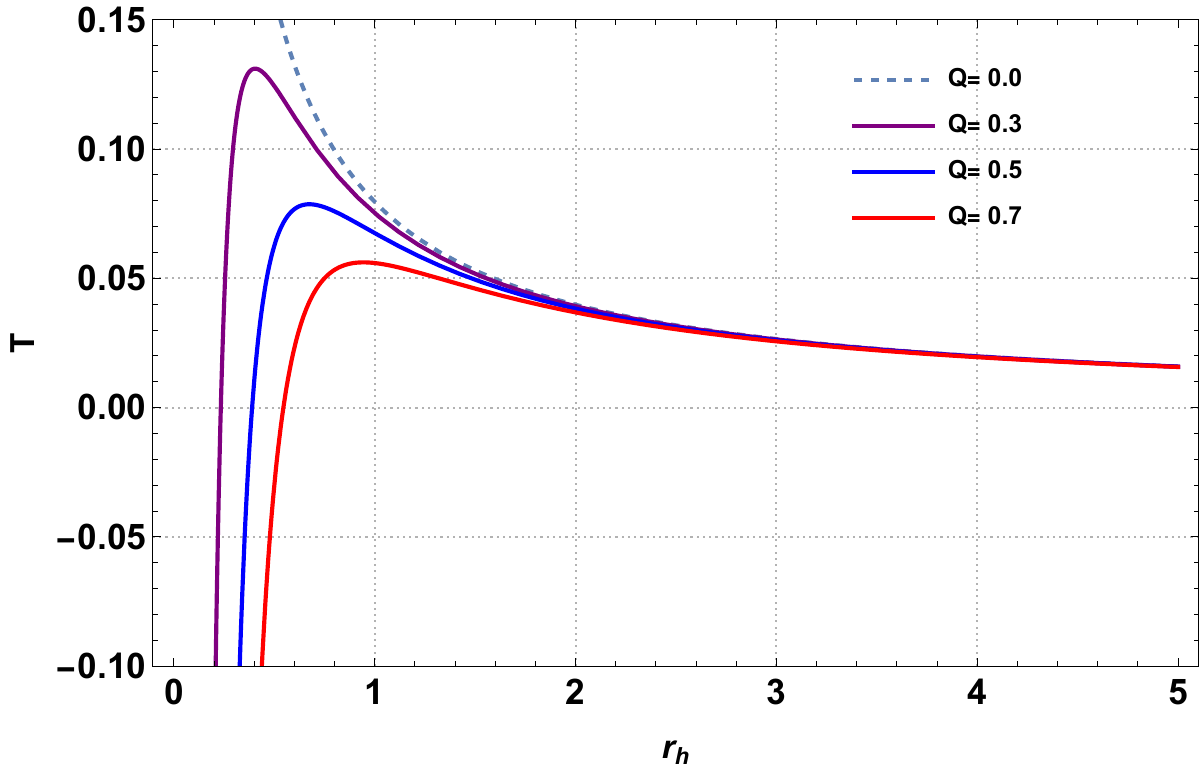}\qquad
    \includegraphics[width=0.45\linewidth]{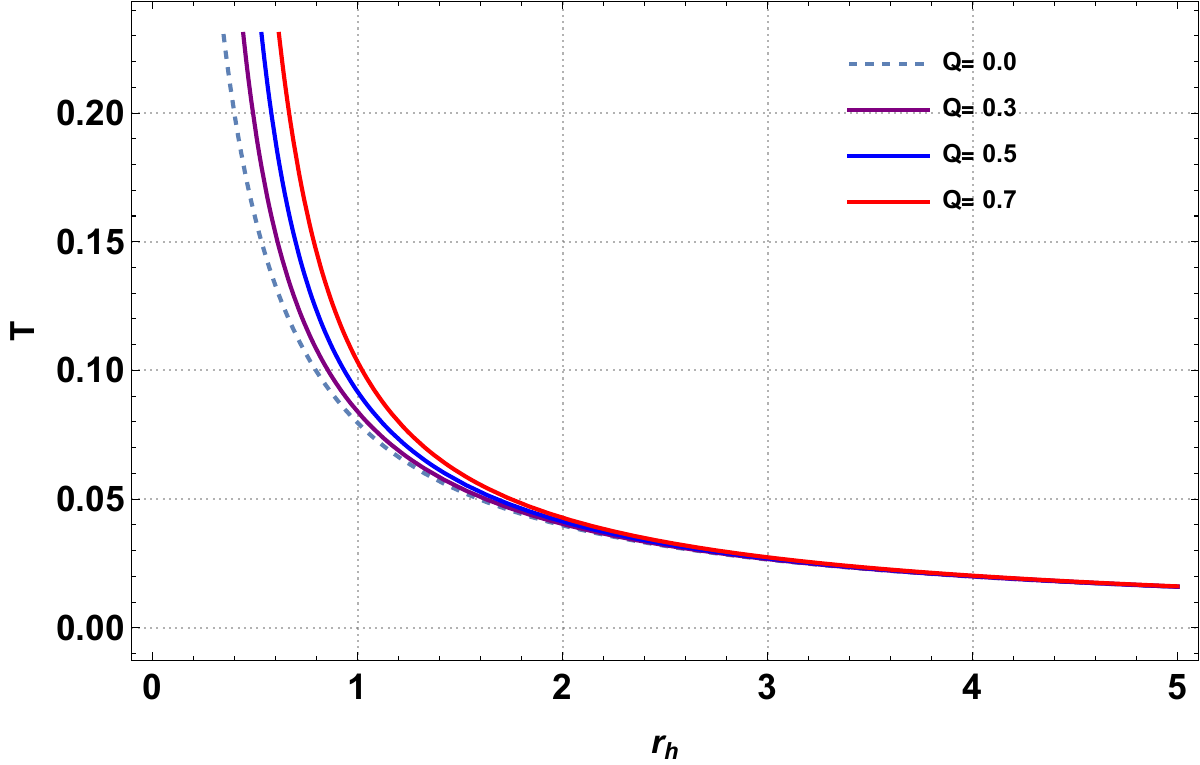}\\
    (i) $\eta=+1$ \hspace{6cm} (ii) $\eta=-1$\\
    \caption{Behavior of the Hawking temperature for Mod(A)Max black hole by varying the electric charge $Q$, while keeping ModMax's parameter fixed $\gamma=0.5$.}
    \label{fig:temperature}
\end{figure}
In Fig.~\ref{fig:temperature}, we illustrate the Hawking temperature as a function of horizon by varying the electric charge for $\eta=+1$ and $\eta=-1$, while keeping ModMax's parameter $\gamma=0.5$ fixed.

Finally, the specific heat capacity is given by
\begin{equation}
    C=\frac{dM}{dT}=-2\pi\,r^2_h\,\left(\frac{r^2_h-\eta\,e^{-\gamma}\,Q^2}{r^2_h-3\,\eta\,e^{-\gamma}\,Q^2}\right).\label{heat}
\end{equation}
The above result reduces to the standard Schwarzschild black hole case when $Q = 0$. Below, we discuss some special cases:
\begin{itemize}
    \item For $\eta = +1$, corresponding to the ModMax black hole, the specific heat reduces to
    \begin{equation}
        C = -2\pi\, r_h^{2}\left(\frac{r_h^{2} - e^{-\gamma} Q^{2}}{r_h^{2} - 3 e^{-\gamma} Q^{2}}\right).
        \label{heat2}
    \end{equation}
    This expression further reduces to the Reissner-Nordström black hole result in the limit $\gamma = 0$.

    \item For $\eta = -1$, corresponding to the phantom ModMax or Mod(A)Max black hole, the specific heat takes the form
    \begin{equation}
        C = -2\pi\, r_h^{2}\left(\frac{r_h^{2} + e^{-\gamma} Q^{2}}{r_h^{2} + 3 e^{-\gamma} Q^{2}}\right).
        \label{heat3}
    \end{equation}
    In this case, the expression reduces to the phantom Reissner-Nordström black hole result for $\gamma = 0$.
\end{itemize}

\begin{figure}[tbhp]
    \centering
    \includegraphics[width=0.45\linewidth]{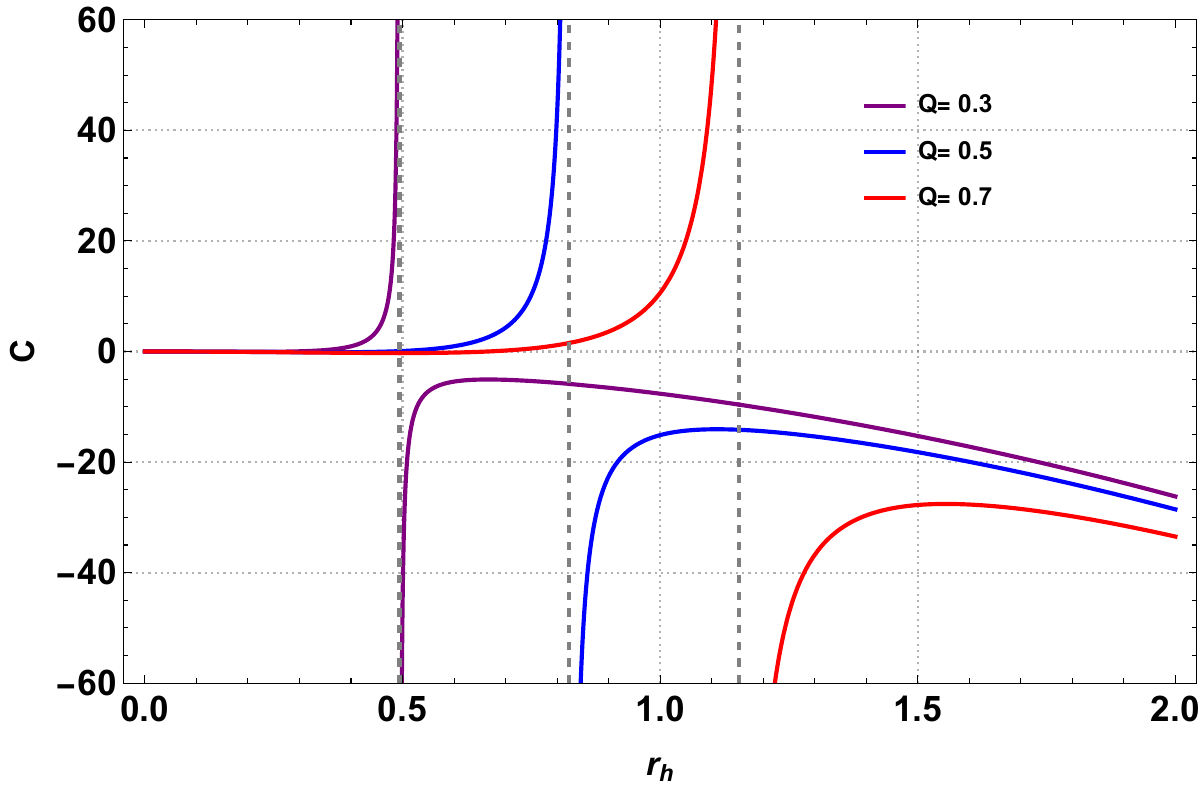}\qquad
    \includegraphics[width=0.45\linewidth]{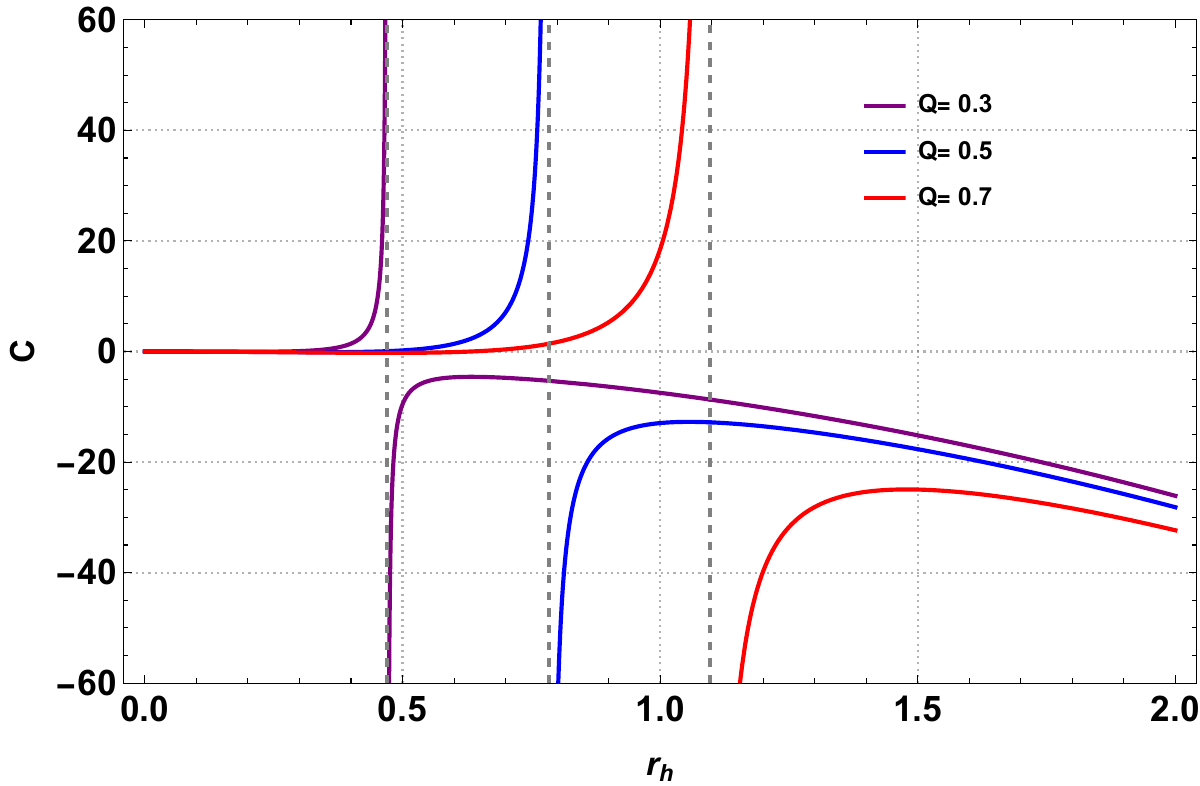}\\
    (i) $\gamma=0.1$ \hspace{6cm} (ii) $\gamma=0.2$\\
    \caption{Behavior of the specific heat capacity for ModMax black hole by varying the electric charge $Q$, while keeping the ModMax's parameter $\eta=+1$.}
    \label{fig:heat}
\end{figure}
In Fig.~\ref{fig:heat}, we illustrate the specific heat capacity as a function of the horizon radius for the ModMax black hole by varying the electric charge \(Q\) and considering two values of the ModMax parameter, \(\gamma = 0.1\) and \(\gamma = 0.2\).

\section{Observable Optical Features of Mod(A)Max BH }\label{sec:3}

In this section, we investigate null geodesics of spacetime using the Lagrangian formalism. Null geodesics play a crucial role in determining several observable features of compact objects, 
including the photon sphere (or photon ring), black hole shadow, and gravitational lensing \cite{FA1,FA2,FA7,FA9,FA11,FA12}. Within this framework, the Lagrangian density describing the motion of massless particles can be written in terms of the spacetime metric tensor $g_{\mu\nu}$ as \cite{SC1983,RMW1984}: 
\begin{equation}
    \mathbb{L}=\frac{1}{2}\,g_{\mu\nu}\, \dot x^{\mu}\,\dot x^{\nu},\label{bb2}
\end{equation}
where dot represents ordinary derivative w. r. to $\lambda$, an affine parameter along geodesics.

The space-time (\ref{metric}) can be expressed as $ds^2=g_{\mu\nu} dx^{\mu} dx^{\nu}$, where the metric tensor $g_{\mu\nu}$ with $\mu,\nu=0,...,3$ is given by
\begin{equation}
    g_{\mu\nu}=\mbox{diag}\left(-f(r),\,\frac{1}{f(r)},\,r^2,\,r^2\, \sin^2 \theta\right),\label{bb1} 
\end{equation}
Considering the geodesic motion in the equatorial plane defined by $\theta=\pi/2$, the Lagrangian density function (\ref{bb2}) using (\ref{bb1}) explicitly can be written as
\begin{equation}
    \mathbb{L}=\frac{1}{2}\left[-f(r) \dot t^2+\frac{1}{f(r)} \dot r^2+r^2\,\dot \phi^2 \right].\label{bb3}
\end{equation}

It is evident that the Lagrangian density function is independent of the temporal coordinate $t$ and the azimuthal angular coordinate $\phi$, corresponding to the Killing vectors $\xi_t^{\mu} \equiv \partial/\partial t$ and $\xi_{\phi}^{\mu} \equiv \partial/\partial \phi$, respectively. Consequently, these coordinates are cyclic, leading to the existence of two conserved quantities along the geodesic motion: the conserved energy $\mathrm{E}$ and the conserved angular momentum $\mathrm{L}$. These quantities are given by
\begin{equation}
    \mathrm{E}=f(r)\,\dot t.\label{bb4}
\end{equation}
And
\begin{equation}
    \mathrm{L}=r^2\,\dot \phi.\label{bb5}
\end{equation}

Based on the above, the radial equation of motion for photon particles ($\mathbb{L}=0$) can be expressed as
\begin{equation}
    \left(\frac{dr}{d\lambda}\right)^2+V_{\rm eff}=\mathrm{E}^2,\label{bb6}
\end{equation}
which is equivalent to the one-dimensional equation of motion of particles, and $V_{\rm eff}$ is the effective potential that governs the photon dynamics. The effective potential is given by
\begin{equation}
    V_{\rm eff}=\frac{\mathrm{L}^2}{r^2}\,f(r)=\frac{\mathrm{L}^2}{r^2}\,\left(1-\frac{2 M}{r}+\eta\,\frac{e^{-\gamma}\,Q^2}{r^2}\right).\label{bb7}
\end{equation}
Noted that in the limit $\eta=+1$ corresponding to the ModMax black hole, the effective potential given in Eq.~(\ref{bb7}) reduces to the result presented in \cite{EGH2024}.

\begin{figure}[ht!]
    \centering
    \includegraphics[width=0.95\linewidth]{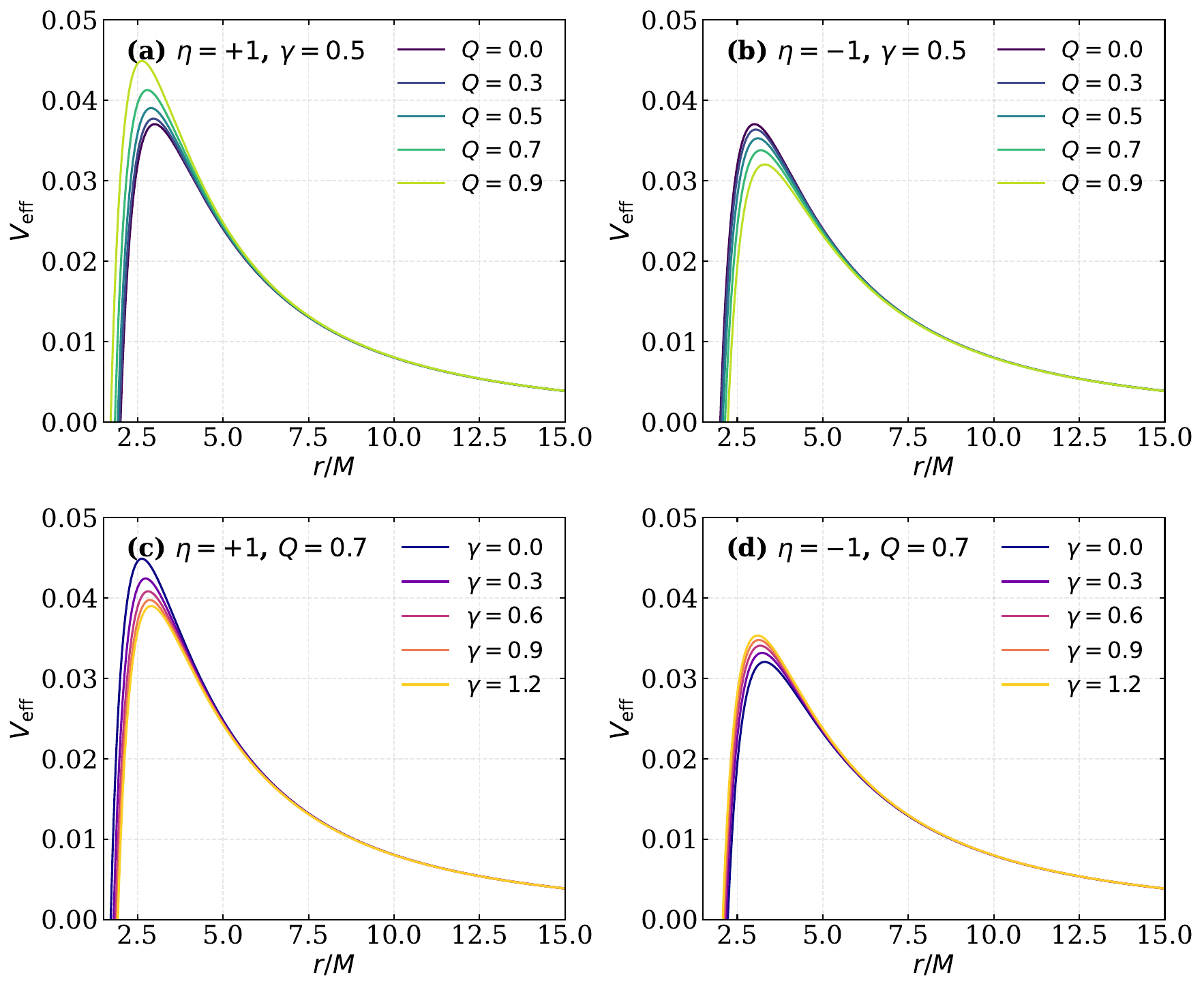}
    \caption{Effective potential $V_{\rm eff}$ for photons as a function of the radial coordinate $r/M$ in the Mod(A)Max black hole spacetime. Figures (a) and (b) show the variation with electric charge $Q$ for fixed $\gamma = 0.5$, while Figs. (c) and (d) display the variation with the ModMax parameter $\gamma$ for fixed $Q = 0.7$. The left column corresponds to the ModMax case ($\eta = +1$) and the right column to the Mod(A)Max case ($\eta = -1$). In all panels, $M = 1$ and $L = 1$.}
    \label{fig:Veff_photon}
\end{figure}

In Fig.~\ref{fig:Veff_photon}, we display the behavior of the effective potential $V_{\rm eff}$ for photons as a function of the radial coordinate for various values of the black hole parameters. Figures \ref{fig:Veff_photon}(a) and \ref{fig:Veff_photon}(b) illustrate the dependence on the electric charge $Q$ for a fixed ModMax parameter $\gamma = 0.5$. In the ModMax case [Fig. \ref{fig:Veff_photon}(a), $\eta = +1$], increasing the charge parameter leads to a decrease in both the height and radial position of the potential barrier maximum. This behavior indicates that electric charge in the ModMax scenario tends to weaken the gravitational trapping of photons, thereby reducing the size of the photon sphere. In contrast, the Mod(A)Max case [Fig. \ref{fig:Veff_photon}(b), $\eta = -1$] exhibits the opposite trend: increasing $Q$ raises the potential barrier and shifts its maximum to larger radii. This enhancement of the potential barrier in the phantom case implies a stronger gravitational influence on photon trajectories, resulting in a larger photon sphere compared to the Schwarzschild limit.

Figures \ref{fig:Veff_photon}(c) and \ref{fig:Veff_photon}(d) of Fig.~\ref{fig:Veff_photon} illustrate the influence of the ModMax coupling parameter $\gamma$ on the photon effective potential for a fixed charge $Q = 0.7$. In the ModMax case [Fig. (c), $\eta = +1$], increasing $\gamma$ leads to a systematic increase in the potential barrier height, suggesting that stronger nonlinear electromagnetic corrections counteract the charge-induced reduction of the photon sphere. For the Mod(A)Max case [Fig. (d), $\eta = -1$], larger values of $\gamma$ suppress the growth of the potential barrier, moderating the enhancement caused by the phantom charge contribution. These results demonstrate that the interplay between the electric charge $Q$ and the ModMax parameter $\gamma$ provides a rich phenomenology for the optical properties of Mod(A)Max black holes, with potentially observable consequences for the photon sphere radius and black hole shadow.

For circular null orbits, the conditions $\frac{dr}{d\lambda}=0$ and $\frac{d^2r}{d\lambda^2}=0$ must be satisfied. From Eq. (\ref{bb6}), we find the following relations:
\begin{equation}
    \mathrm{E}^2=V_{\rm eff}=\frac{\mathrm{L}^2}{r^2}\,f(r).\label{bb8}
\end{equation}
And
\begin{equation}
    \frac{d}{dr}\left(\frac{f(r)}{r^2}\right)=0\Rightarrow 2\,r\,f(r)-r^2\,f'(r)=0.\label{bb9}
\end{equation}

Now, we aim to determine the photon sphere radius of the black hole. The photon sphere is a region around a black hole where photons can orbit the black hole on unstable circular orbits due to strong gravitational lensing. It plays a central role in determining optical observables such as the black hole shadow, light bending, and gravitational lensing. 

In our case at hand, the relation (\ref{bb9}) will give us the photon sphere radius $r=r_s$. Simplification of (\ref{bb9}) results 
\begin{align}
r^2-3 M r+2\eta\,e^{-\gamma}\,Q^2= 0.\label{bb10}
\end{align}
The exact analytical solution of the above equation is given by
\begin{equation}
r_s=\frac{3 M}{2}\left[1+\sqrt{1-\frac{8 \eta e^{-\gamma}\,Q^2}{9 M^2}}\right].\label{bb11}
\end{equation}
From the photon sphere radius (\ref{bb11}), we observe that the geometric parameters, such as the black hole mass $M$, the electric charge $Q$, and the ModMax parameter $\gamma$, significantly influence the photon sphere size. Moreover, depending on the sign of $\eta$, this radius can be either larger or smaller compared to the Schwarzschild case ($r^{\rm Sch}_s=3 M$).

\begin{itemize}
    \item For $\eta=1$, corresponding to the ModMax charged black hole, the photon sphere radius $r_s$ is given by \cite{EGH2024}
\begin{equation}
r_s=\frac{3 M}{2}\left[1+\sqrt{1-\frac{8e^{-\gamma}\,Q^2}{9 M^2}}\right]<r^{\rm Sch.}_{s}.\label{bb11a}
\end{equation}
Noted that the photon sphere exists, provided we have the following constraint on the parameters
\begin{equation}
    9 M^2 >8 e^{-\gamma}\,Q^2.\label{bb11b}
\end{equation}

\item For $\eta=-1$, corresponding to the phantom ModMax or Mod(A)Max black hole, the photon sphere radius is given by
\begin{equation}
r_s=\frac{3 M}{2}\left[1+\sqrt{1+\frac{8 e^{-\gamma}\,Q^2}{9 M^2}}\right]>r^{\rm Sch.}_{s}.\label{bb11c}
\end{equation}
\end{itemize}

\begin{figure}[ht!]
\centering
\includegraphics[width=0.95\linewidth]{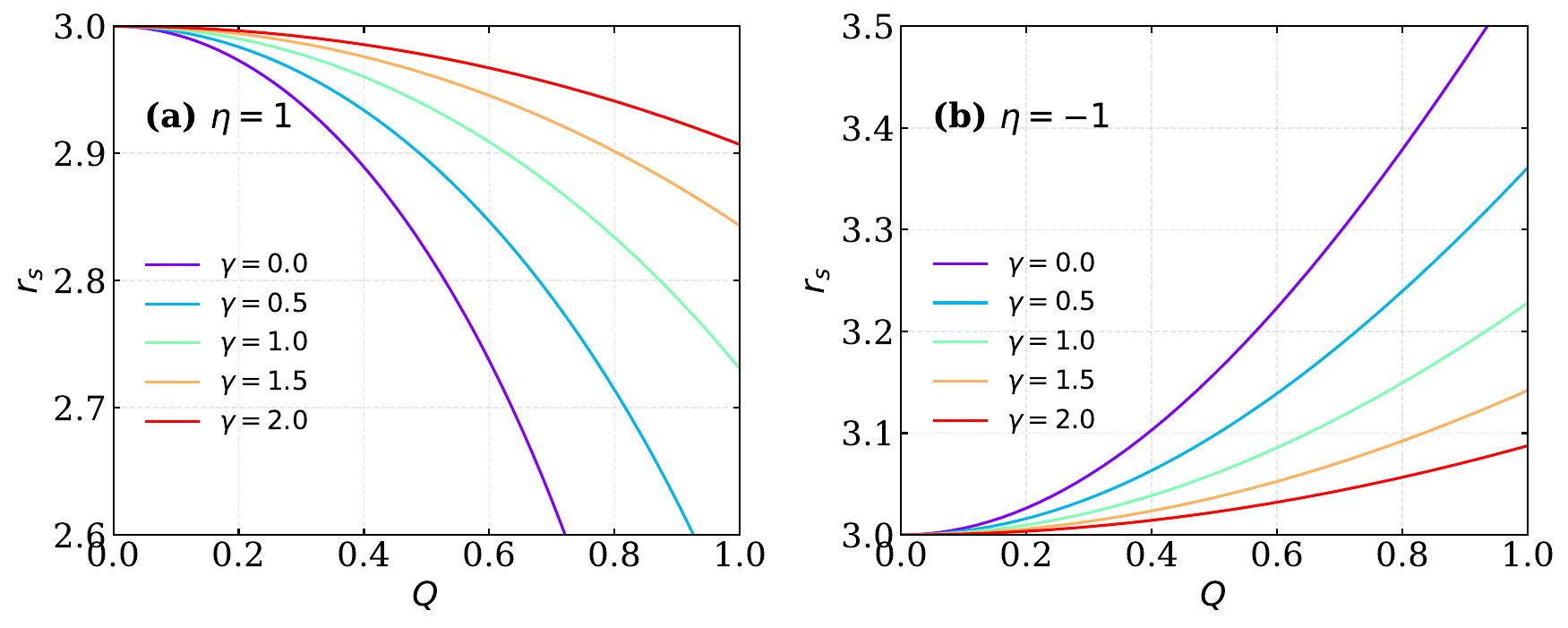}
\caption{Behavior of the photon sphere radius $r_s$ for a Mod(A)Max black hole as a function of the charge parameter $Q$. Panel (a) illustrates the ModMax case ($\eta = +1$), while panel (b) shows the Mod(A)Max case ($\eta = -1$). Different curves correspond to different values of the ModMax parameter $\gamma$. Here $M = 1$.}
\label{fig:photon-radius}
\end{figure}

In Fig.~\ref{fig:photon-radius}, we display how the photon sphere radius $r_s$ varies with the charge parameter $Q$ for both ModMax and Mod(A)Max black holes. The photon sphere is a region surrounding the black hole where photons can move along unstable circular orbits due to strong gravitational lensing. It plays a central role in determining optical observables such as the black hole shadow, light bending, and gravitational lensing.

Figure \ref{fig:photon-radius}(a) illustrates the ModMax case ($\eta = +1$). As the charge parameter $Q$ increases, the photon sphere radius $r_s$ decreases, showing that the presence of charge tends to pull the photon orbit closer to the black hole. Increasing the ModMax parameter $\gamma$ leads to a systematic increase in $r_s$, suggesting that nonlinear electromagnetic corrections counteract the attractive effect of the charge. In contrast, Fig. \ref{fig:photon-radius}(b) shows the Mod(A)Max case ($\eta = -1$), where the photon sphere radius $r_s$ increases with increasing charge $Q$, revealing a qualitatively different influence of the electromagnetic sector. Larger values of $\gamma$ further diminish this growth. These results demonstrate that, depending on the sign of $\eta$, the photon sphere radius can be either smaller (for $\eta = +1$) or larger (for $\eta = -1$) compared to the Schwarzschild value $r_s^{\rm Sch} = 3M$.

Next, we determine the shadow radius of the selected black hole and analyze the results. The shadow radius of a black hole corresponds to the apparent size of the dark region seen by a distant observer, caused by photons captured by the black hole. It is directly determined by the photon sphere and the spacetime geometry. Recent advancements, such as the images of the supermassive black holes at the centers of M87$^{\ast}$~\cite{EHTL1,EHTL4,EHTL5,EHTL6} and Sgr A$^{\ast}$~\cite{EHTL12,EHTL14,EHTL15,EHTL16,EHTL17} captured by the Event Horizon Telescope (EHT), have ushered in a new era of black hole astronomy. These groundbreaking observations not only directly confirm the existence of black holes but also impose stringent observational constraints on theories of gravity in the strong-field regime.

As the lapse function is asymptotically flat space at radial infinity, that is, $\lim_{r \to \infty} f(r) =1$, the shadow radius for a static distant observer is given by \cite{Volker2022}
\begin{equation}
   R_{\rm sh}=\frac{r_s}{\sqrt{f(r_s)}}=\beta_{\rm ph}.\label{bb13}
\end{equation}
Substituting the metric function $f(r)$ and the photon sphere radius $r_s$, we find the shadow radius as follows: 
\begin{equation}
   R_{\rm sh}=\frac{r_s}{\sqrt{1-\frac{2 M}{r_s}+\eta\,\frac{e^{-\gamma}\,Q^2}{r^2_s}}}=\frac{3\sqrt{3}M}{2\sqrt{2}}\,\frac{\left(1+\sqrt{1-\frac{8 \eta e^{-\gamma}\,Q^2}{9 M^2}}\right)^2}{\sqrt{1+\sqrt{1-\frac{8 \eta e^{-\gamma}\,Q^2}{9 M^2}}-\frac{2 \eta e^{-\gamma} Q^2}{3 M^2}}}.\label{bb14}
\end{equation}
Noted that photons with impact parameter $\beta (=\mathrm{L}/\mathrm{E})<\beta_{\rm ph}$ are captured by the black hole. Photons with $\beta>\beta_{\rm ph}$ can escape to infinity, forming the bright surrounding region. 

Analogue to the photon sphere, the shadow radius gets modified by the black hole mass $M$, the electric charge $Q$, and the ModMax parameter $\gamma$. Moreover, depending on the sign of $\eta$, this radius can be either larger or smaller compared to the Schwarzschild case ($R^{\rm Sch}_{\rm sh}=3\sqrt{3} M$).
\begin{itemize}
    \item For $\eta=1$, corresponding to the ModMax black hole, the shadow radius is given by
\begin{equation}
R_{\rm sh}=\frac{3\sqrt{3}M}{2\sqrt{2}}\,\frac{\left(1+\sqrt{1-\frac{8e^{-\gamma}\,Q^2}{9 M^2}}\right)^2}{\sqrt{1+\sqrt{1-\frac{8 e^{-\gamma}\,Q^2}{9 M^2}}-\frac{2 e^{-\gamma} Q^2}{3 M^2}}}.\label{bb14a}
\end{equation}

\item For $\eta=-1$, corresponding to the phantom ModMax or Mod(A)Max black hole, the shadow radius is given by
\begin{equation}
R_{\rm sh}=\frac{3\sqrt{3}M}{2\sqrt{2}}\,\frac{\left(1+\sqrt{1+\frac{8e^{-\gamma}\,Q^2}{9 M^2}}\right)^2}{\sqrt{1+\sqrt{1+\frac{8 e^{-\gamma}\,Q^2}{9 M^2}}+\frac{2 e^{-\gamma} Q^2}{3 M^2}}}.\label{bb14b}
\end{equation}
\end{itemize}

\begin{figure}[ht!]
\centering
\includegraphics[width=0.95\linewidth]{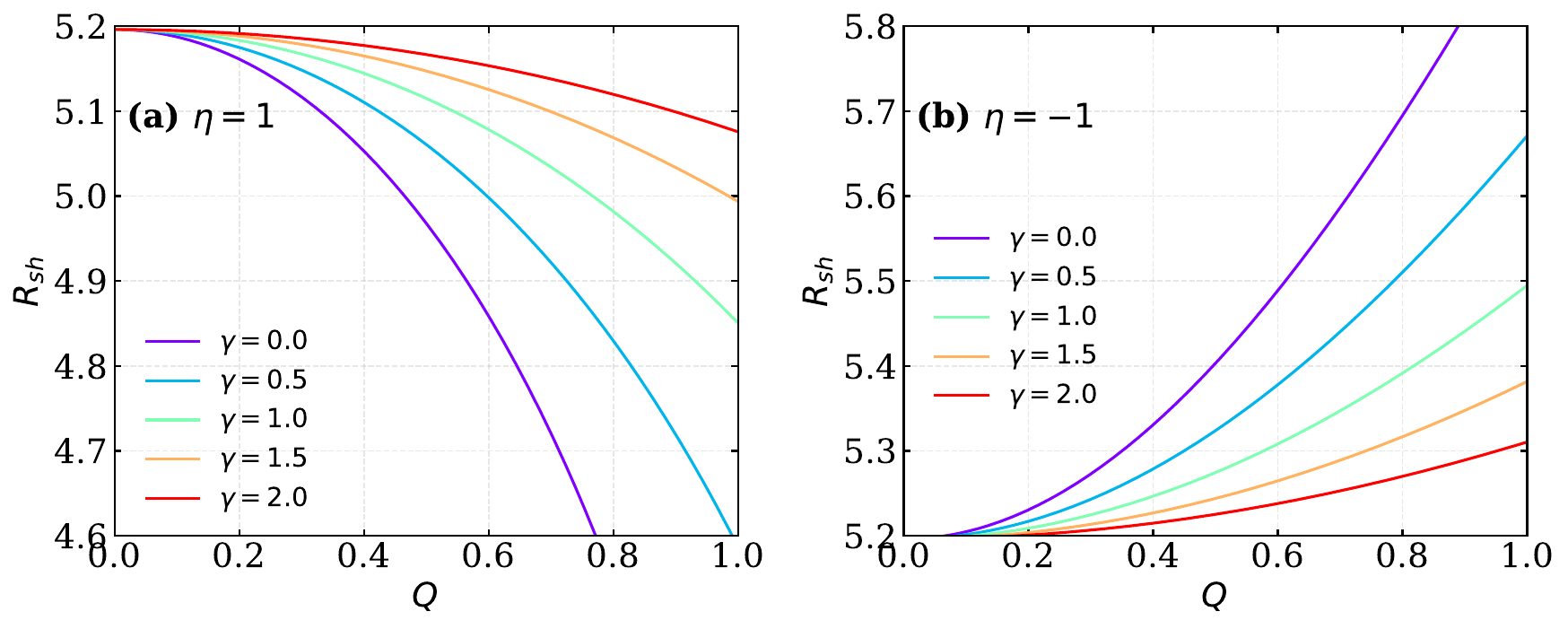}
\caption{Behavior of the shadow radius $R_{sh}$ for a Mod(A)Max black hole as a function of the charge parameter $Q$. Panel (a) illustrates the ModMax case ($\eta = +1$), while panel (b) shows the Mod(A)Max case ($\eta = -1$). Different curves correspond to different values of the ModMax parameter $\gamma$. Here $M = 1$.}
\label{fig:shadow-radius}
\end{figure}

In Fig.~\ref{fig:shadow-radius}, we display how the shadow radius $R_{sh}$ varies with the charge parameter $Q$ for both ModMax and Mod(A)Max black holes. The shadow radius corresponds to the apparent size of the dark region seen by a distant observer, caused by photons captured by the black hole. It is directly determined by the photon sphere and the spacetime geometry, and recent observations by the Event Horizon Telescope (EHT) of the supermassive black holes M87* and Sgr A* have provided direct measurements of this quantity, offering stringent constraints on theories of gravity in the strong-field regime.

According to the observational datat by EHT collaboration~\cite{EHTL1,EHTL12}, the reported classical shadow angular diameter is \(\theta_{\mathrm{M87}^*} = 42 \pm 3\,\mu\mathrm{as}\). Other parameters include the distance of M87* from the Earth, \(D = 16.8\,\mathrm{Mpc}\), and the mass of M87*, \(M_{\mathrm{M87}^*} = 6.5 \pm 0.90 \times 10^{9} M_{\odot}\). For Sgr. A*, the angular shadow diameter is \(\theta_{\mathrm{Sgr.\,A}^*} = 48.7 \pm 7\,\mu\mathrm{as}\) (EHT), the distance from the Earth is \(D = 8277 \pm 33\,\mathrm{pc}\), and the mass of the black hole is \(M_{\mathrm{Sgr.\,A}^*} = 4.3 \pm 0.013 \times 10^{6} M_{\odot}\) (VLTI) \cite{ss20}. The respective diameter of the shadow size in units of the black hole mass can be determined using
\( d_{\mathrm{sh}} = \frac{D \theta}{M},\). Therefore, the theoretical shadow diameter can be obtained via \(d_{\mathrm{sh}}^{\mathrm{theo}} = 2 \mathcal{R}_{\mathrm{sh}}\). Therefore, the diameter of the shadow image of M87* and Sgr A* as \( d_{\mathrm{sh}}^{\mathrm{M87}^*} = (11 \pm 1.5) M, \quad \text{and} \quad d_{\mathrm{sh}}^{\mathrm{Sgr.\,A}^*} = (9.5 \pm 1.4) M,\) respectively.

Figure \ref{fig:shadow-radius}(a) illustrates the ModMax case ($\eta = +1$). As the charge parameter $Q$ increases, the shadow radius $R_{sh}$ decreases. Increasing the ModMax parameter $\gamma$ leads to a systematic increase in $R_{sh}$, counteracting the charge-induced reduction. In contrast, Fig. \ref{fig:shadow-radius}(b) shows the Mod(A)Max case ($\eta = -1$), where the shadow radius $R_{sh}$ increases with increasing charge $Q$, whereas larger values of $\gamma$ further diminish this growth. Analogous to the photon sphere, the shadow radius is modified by the black hole mass $M$, the electric charge $Q$, and the ModMax parameter $\gamma$. Depending on the sign of $\eta$, the shadow radius can be either smaller (for $\eta = +1$) or larger (for $\eta = -1$) compared to the Schwarzschild value $R_{sh}^{\rm Sch} = 3\sqrt{3}M \approx 5.196M$. These distinct behaviors provide potential observational signatures that could be used to constrain the parameters of Mod(A)Max black holes through future high-precision shadow measurements.

Finally, we determine the photon trajectories and analyze how various parameters alter their trajectory. The equation of orbit using Eqs. (\ref{bb6}) and (\ref{bb7}) is defined as,
\begin{equation}
    \left(\frac{d\phi}{dr}\right)^2=\frac{\dot \phi^2}{\dot r^2}=r^4\,\left[\frac{1}{\beta^2}-\frac{1}{r^2}\,\left(1-\frac{2 M}{r}+\eta\,\frac{e^{-\gamma}\,Q^2}{r^2}\right)\right].\label{bb17}
\end{equation}
Transforming to anew variable via $r(\phi)=\frac{1}{u(\phi)}$ and after simplification results
\begin{equation}
    \left(\frac{du}{d\phi}\right)^2+u^2=2 M u^3-\eta e^{-\gamma}\,Q^2 u^4.\label{bb18}
\end{equation}
Differentiating both sides w. r. to $\phi$ and after simplification results
\begin{equation}
    \frac{d^2u}{d\phi^2}+u=3 M u^2-2\eta e^{-\gamma}\,Q^2 u^3.\label{bb19}
\end{equation}

Equation (\ref{bb19}) is a second-order homogeneous differential equation representing photon trajectories in the gravitational field produced by Mod(A)Max black hole. Its perturbative solution assuming $u(\phi)=u_0(\phi)+\varepsilon u_1(\phi)+\mathcal{O}(\varepsilon^2)$ with $\varepsilon \ll 1$ is given by
\begin{equation}
u(\phi) = \frac{1}{\beta}\,\cos\phi+\varepsilon\left[\frac{3 M}{2 \beta^2} 
- \frac{M}{2 \beta^2} \cos 2\phi- \frac{3 \eta e^{-\gamma} Q^2}{4 \beta^3}\, \phi \sin\phi
+ \frac{\eta e^{-\gamma} Q^2}{16 \beta^3} \cos 3\phi\right]+\mathcal{O}(\varepsilon^2) \,.\label{bb20}
\end{equation}
Here, the first term corresponds to the zeroth-order straight-line photon trajectory, while the remaining terms represent first-order corrections due to the black hole mass $M$ and the charge term $Q^2$. The secular term $-(3 \eta e^{-\gamma} Q^2 /4 \beta^3)\, \phi \sin\phi$ is responsible for the gravitational deflection of light.

The total gravitational deflection angle of photons in the weak field limit (up to first-order term) is given by
\begin{equation}
\Delta \phi \simeq \frac{4 M}{\beta} - \frac{3 \pi \eta e^{-\gamma} Q^2}{4 \beta^2},\label{bb21}
\end{equation}
where the first term corresponds to the usual Schwarzschild bending and the second term represents the contribution from the electric charge $Q$ and the Mod(A)Max's parameter $\eta e^{-\gamma}$. This result follows directly from the secular term in the perturbative solution of the photon trajectory equation. It is worth noting that
there is no $\pi$ in the second term because we used the secular term approximation (linear in $\phi$)-not the full integral. It is evident that the deflection angle is less in the ModMax black hole compared to the phantom ModMax or Mod(A)Max black hole case, that is, $\Delta \phi(\eta=+1) < \Delta \phi(\eta=-1)$.

In the limit $\eta = +1$ with vanishing ModMax parameter $\gamma = 0$, the spacetime reduces to the standard Reissner-Nordstr\"om black hole. In this case, the deflection angle reduces to
\begin{equation}
\Delta \phi_{\rm RN} \simeq \frac{4 M}{\beta} - \frac{3 \pi Q^2}{4 \beta^2}, \label{bb22}
\end{equation}
which agrees with the results reported in \cite{Virbhadra2000,Eiroa2002}.

On the other hand, for $\eta = -1$ and $\gamma = 0$, the spacetime corresponds to a phantom Reissner-Nordstr\"om black hole. The corresponding deflection angle is then
\begin{equation}
\Delta \phi_{\rm phantom\,RN} \simeq \frac{4 M}{\beta} + \frac{3 \pi Q^2}{4 \beta^2}. \label{bb23}
\end{equation}

\begin{figure}[ht!]
    \centering
    \includegraphics[width=0.95\linewidth]{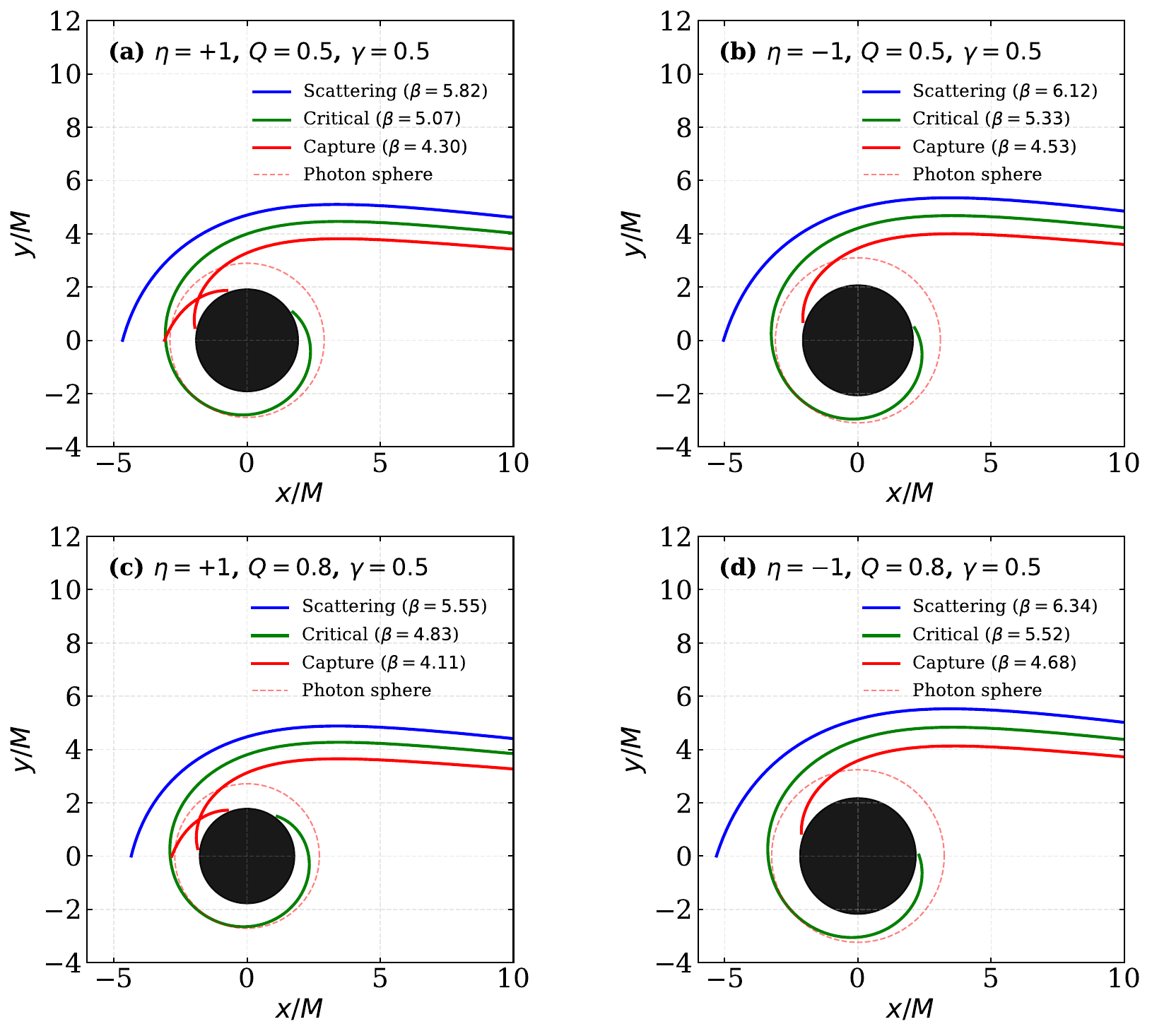}
    \caption{Photon trajectories around the Mod(A)Max black hole for three different regimes: scattering (blue), critical (green), and capture (red). The black-filled circle represents the event horizon, and the red-dashed circle indicates and determines the sphere. Panels (a) and (b) correspond to $Q = 0.5$, while panels (c) and (d) show $Q = 0.8$. The left column displays the ModMax case ($\eta = +1$) and the right column the Mod(A)Max case ($\eta = -1$). In all panels, $M = 1$ and $\gamma = 0.5$. The critical impact parameter $\beta_c$ varies with the spacetime parameters, determining the boundary between scattering and capture orbits.}
    \label{fig:photon_trajectory}
\end{figure}

In Fig.~\ref{fig:photon_trajectory}, we display the photon trajectories around the Mod(A)Max black hole for three distinct dynamical regimes. The scattering regime (blue curves) corresponds to photons with impact parameter $\beta > \beta_c$, which approach the black hole, experience gravitational deflection, and escape to infinity. The critical regime (green curves) represents photons with $\beta \approx \beta_c$, which orbit multiple times near the photon sphere before eventually escaping or falling into the black hole. The capture regime (red curves) corresponds to $\beta < \beta_c$, where photons cross the photon sphere and inevitably fall into the event horizon.

The photon sphere, indicated by the red dashed circle, plays a central role in determining the optical properties of the black hole. Comparing panels (a) and (b) for $Q = 0.5$, we observe that the critical impact parameter is smaller in the ModMax case ($\eta = +1$) than in the Mod(A)Max case ($\eta = -1$), consistent with the behavior of the photon sphere radius discussed previously. This difference becomes more pronounced for larger charge values, as seen in panels (c) and (d) for $Q = 0.8$. In the ModMax case [panel (c)], the reduced photon sphere leads to tighter critical orbits, while in the Mod(A)Max case [panel (d)], the enlarged photon sphere results in wider critical trajectories.

These results demonstrate that the interplay between the electric charge $Q$ and the parameter $\eta$ significantly affects the photon dynamics around the black hole. The critical impact parameter, which determines the boundary between scattering and capture, is directly related to the shadow radius observed by distant observers. Therefore, the distinct photon trajectories in the ModMax and Mod(A)Max scenarios could, in principle, lead to observable differences in the black hole shadow and gravitational lensing signatures.

\section{Dynamics of Neutral Test Particles}\label{sec:4}

In this section, we investigate the dynamics of neutral test particles around a phantom ModMax black hole. We derive the effective potential and analyze how the spacetime parameters influence the specific energy, specific angular momentum, and ISCO location for particles in circular orbits.

The motion of a neutral particle with mass $m$ is governed by the Hamiltonian:
\begin{equation}
    H = \frac{1}{2}g^{\mu\nu}p_{\mu}p_{\nu} + \frac{1}{2}m^2,\label{cc1}
\end{equation}
where $p^{\mu} = mu^{\mu}$ is the 4-momentum, $u^{\mu} = dx^{\mu}/d\tau$ is the 4-velocity, and $\tau$ is the proper time. The Hamiltonian equations of motion are
\begin{equation}
\frac{dx^\mu}{d\zeta} \equiv mu^\mu = \frac{\partial H}{\partial p_\mu}, \qquad 
\frac{dp_\mu}{d\zeta} = -\frac{\partial H}{\partial x^\mu},\label{cc2}
\end{equation}
where $\zeta = \tau/m$ is the affine parameter.

The static, spherically symmetric spacetime admits two Killing vectors associated with time translation and axial rotation, yielding two conserved quantities:
\begin{equation}
\frac{p_t}{m} = -f(r)\frac{dt}{d\tau} = -\mathcal{E},\label{cc3}
\end{equation}
\begin{equation}
\frac{p_\phi}{m} = r^2\sin^2\theta\frac{d\phi}{d\tau} = \mathcal{L},\label{cc4}
\end{equation}
where $\mathcal{E}$ and $\mathcal{L}$ are the specific energy and specific angular momentum (per unit mass), respectively.

The 4-velocity components are:
\begin{align}
    &\frac{dt}{d\tau} = \frac{\mathcal{E}}{f(r)},\label{cc5}\\
    &\frac{d\phi}{d\tau} = \frac{\mathcal{L}}{r^2\sin^2\theta},\label{cc6}\\
    &\frac{d\theta}{d\tau} = \frac{p_{\theta}}{mr^2},\label{cc7}\\
    &\frac{dr}{d\tau} = \sqrt{\mathcal{E}^2 - \left(\epsilon + \frac{\mathcal{L}^2}{r^2\sin^2\theta} + \frac{p^2_{\theta}}{mr^2}\right)f(r)},\label{cc8}
\end{align}
where $\epsilon = 1$ for time-like particles and $\epsilon = 0$ for light-like particles.

In our work, we focus only on time-like
particles. Therefore, for time-like particles, the Hamiltonian in Eq.~\eqref{cc1} can be rewritten as:
\begin{equation}
    H = \frac{f(r)}{2}p^2_r + \frac{p^2_{\theta}}{r^2} + \frac{1}{2}\frac{m^2}{f(r)}\left[U_{\rm eff}(r,\theta) - \mathcal{E}^2\right],\label{cc9}
\end{equation}
where the effective potential $U_{\rm eff}(r,\theta)$ takes the form:
\begin{equation}
    U_{\rm eff}(r,\theta) = \left(1 + \frac{\mathcal{L}^2}{r^2\sin^2\theta}\right)f(r)=\left(1 + \frac{\mathcal{L}^2}{r^2\sin^2\theta}\right)\,\left(1-\frac{2 M}{r}+\eta\,\frac{e^{-\gamma}\,Q^2}{r^2}\right).\label{cc10}
\end{equation}

In the equatorial plane ($\theta = \pi/2$), the effective potential becomes:
\begin{equation}
    U_{\rm eff}(r) = \left(1 + \frac{\mathcal{L}^2}{r^2}\right)\left[1-\frac{2 M}{r}+\eta\,\frac{e^{-\gamma}\,Q^2}{r^2}\right].\label{cc11}
\end{equation}

\begin{figure}[ht!]
    \centering
    \includegraphics[width=0.95\linewidth]{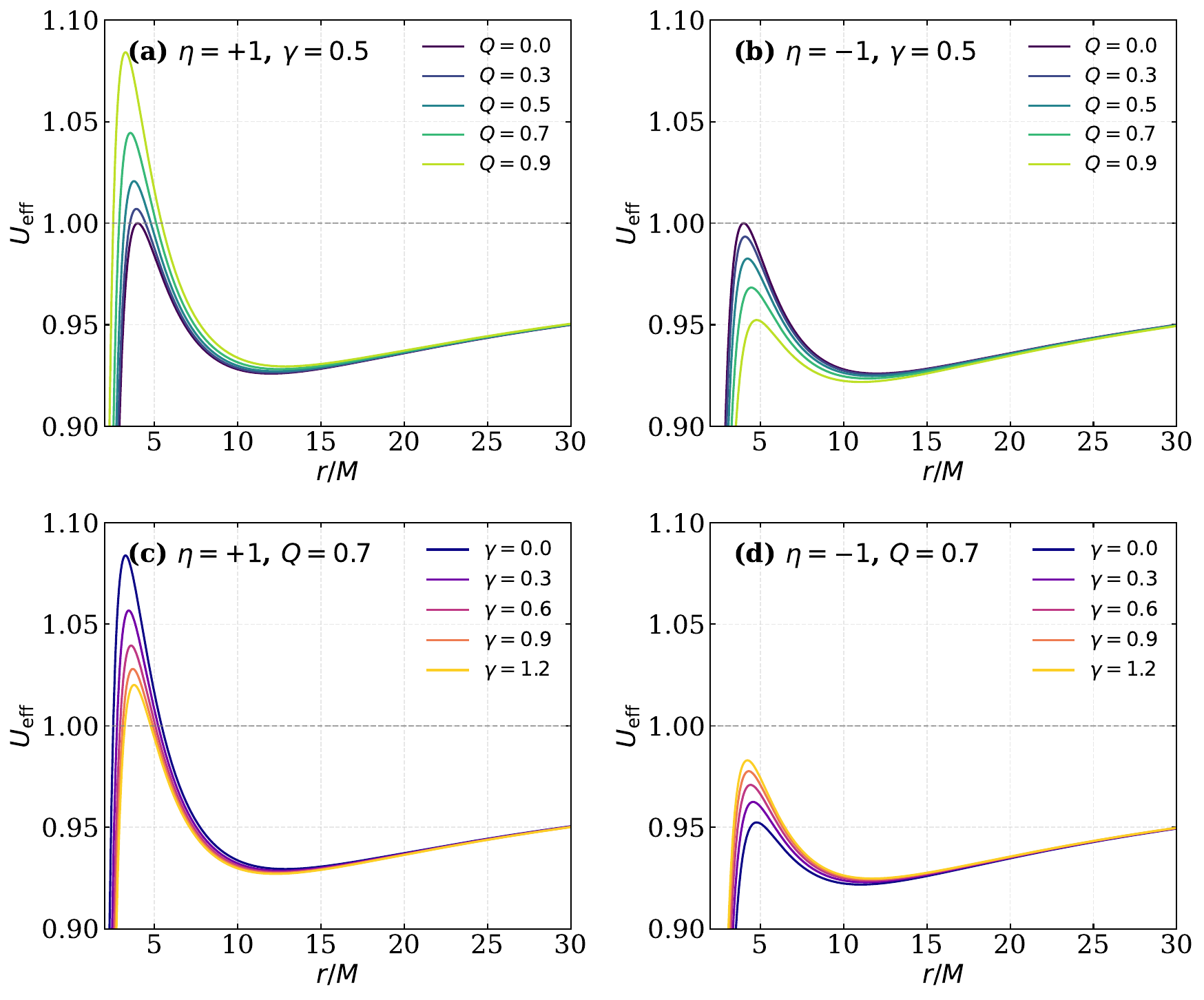}
    \caption{Effective potential $U_{\rm eff}$ for massive test particles as a function of the radial coordinate $r/M$ in the Mod(A)Max black hole spacetime. Figures (a) and (b) show the variation with electric charge $Q$ for fixed $\gamma = 0.5$, while Figs. (c) and (d) display the variation with the ModMax parameter $\gamma$ for fixed $Q = 0.7$. The left column corresponds to the ModMax case ($\eta = +1$) and the right column to the Mod(A)Max case ($\eta = -1$). The horizontal dashed gray line indicates $U_{\rm eff} = 1$, corresponding to a particle at rest at infinity. In all panels, $M = 1$ and $\mathcal{L} = 4$.}
    \label{fig:Ueff_massive}
\end{figure}

In Fig.~\ref{fig:Ueff_massive}, we present the effective potential $U_{\rm eff}$ for massive test particles moving in the equatorial plane of the Mod(A)Max black hole. The effective potential exhibits the characteristic shape expected for black hole spacetimes: a local minimum corresponding to stable circular orbits and, for sufficiently high angular momentum, a local maximum corresponding to unstable circular orbits. The horizontal dashed line at $U_{\rm eff} = 1$ represents the energy of a particle at rest at spatial infinity, serving as a reference for bound and unbound orbits.

Figures \ref{fig:Ueff_massive}(a) and \ref{fig:Ueff_massive}(b) illustrate the dependence of the effective potential on the electric charge $Q$ for a fixed ModMax parameter $\gamma = 0.5$. In the ModMax case [Fig. \ref{fig:Ueff_massive}(a), $\eta = +1$], increasing the charge $Q$ raises the potential barrier and shifts the location of the local minimum to smaller radii. This behavior suggests that the electric charge in the ModMax scenario enhances the gravitational binding, allowing stable circular orbits closer to the black hole. Conversely, in the Mod(A)Max case [Fig. \ref{fig:Ueff_massive}(b), $\eta = -1$], increasing $Q$ lowers the potential barrier and moves the minimum to larger radii, indicating a weakening of the effective gravitational attraction due to the phantom charge contribution.

Figures \ref{fig:Ueff_massive}(c) and \ref{fig:Ueff_massive}(d) show the influence of the ModMax coupling parameter $\gamma$ on the effective potential for a fixed charge $Q = 0.7$. In the ModMax case [Fig. \ref{fig:Ueff_massive}(c), $\eta = +1$], increasing $\gamma$ systematically lowers the potential barrier, counteracting the charge-induced enhancement observed in panel (a). For the Mod(A)Max case [Fig. \ref{fig:Ueff_massive}(d), $\eta = -1$], larger values of $\gamma$ raise the potential barrier, partially compensating the weakening effect of the phantom charge. These results demonstrate that the parameters $Q$ and $\gamma$ have competing effects on the orbital dynamics, with their interplay determining the location and stability of circular orbits around the Mod(A)Max black hole.
The effective potential depends on all spacetime parameters: the electric charge $Q$, and the BH mass $M$, and ModMax's parameter $\gamma$. Additionally, the specific angular momentum $\mathcal{L}$ modifies the centrifugal barrier contribution. The effective potential $U_{\rm eff}(r, \theta)$ is crucial for understanding the motion of the test particles, as it allows one to describe the trajectories of the particles without directly solving the equations of motion.

Next, we determine the effective force experienced by the test particles. The effective force acting on  test particles is defined as the negative gradient of the effective potential $U_{\rm eff} (r, \theta)$ and is given by 
\begin{equation}
    \mathcal{F} = -\frac{1}{2}\frac{\partial U_{\rm eff}}{\partial r}.\label{force1}
\end{equation}
Substituting the effective potential given in Eq.~\eqref{cc10} and simplifying yields:
\begin{equation}
\mathcal{F}(r,\theta) = -\frac{f'(r)}{2} + \frac{\mathcal{L}^2}{r^3\sin^2\theta}\left[2f(r) - rf'(r)\right].\label{force2}
\end{equation}

On the equatorial plane ($\theta=\pi$), the effective radial force simplifies as
\begin{equation}
\mathcal{F}(r) =-\frac{M}{r^2}+\eta\,\frac{e^{-\gamma}\,Q^2}{r^3}+\frac{\mathcal{L}^2}{r^3}\left(1-\frac{3 M}{r}+\frac{2 \eta e^{-\gamma}\,Q^2}{r^2}\right).\label{force3}
\end{equation}

\begin{itemize}
    \item When $\eta=+1$, ModMax black hole case, the effective radial force simplifies as
    \begin{equation}
\mathcal{F}(r) =-\frac{M}{r^2}+\frac{e^{-\gamma}\,Q^2}{r^3}+\frac{\mathcal{L}^2}{r^3}\left(1-\frac{3 M}{r}+\frac{2 e^{-\gamma}\,Q^2}{r^2}\right).\label{force4}
\end{equation}

\item When $\eta=-1$, Mod(A)Max or phantom ModMax black hole, the effective radial force simplifies as,
\begin{equation}
\mathcal{F}(r) =-\frac{M}{r^2}-\frac{e^{-\gamma}\,Q^2}{r^3}+\frac{\mathcal{L}^2}{r^3}\left(1-\frac{3 M}{r}-\frac{2 e^{-\gamma}\,Q^2}{r^2}\right).\label{force5}
\end{equation}
\end{itemize}

\begin{figure}[ht!]
    \centering
    \includegraphics[width=0.95\linewidth]{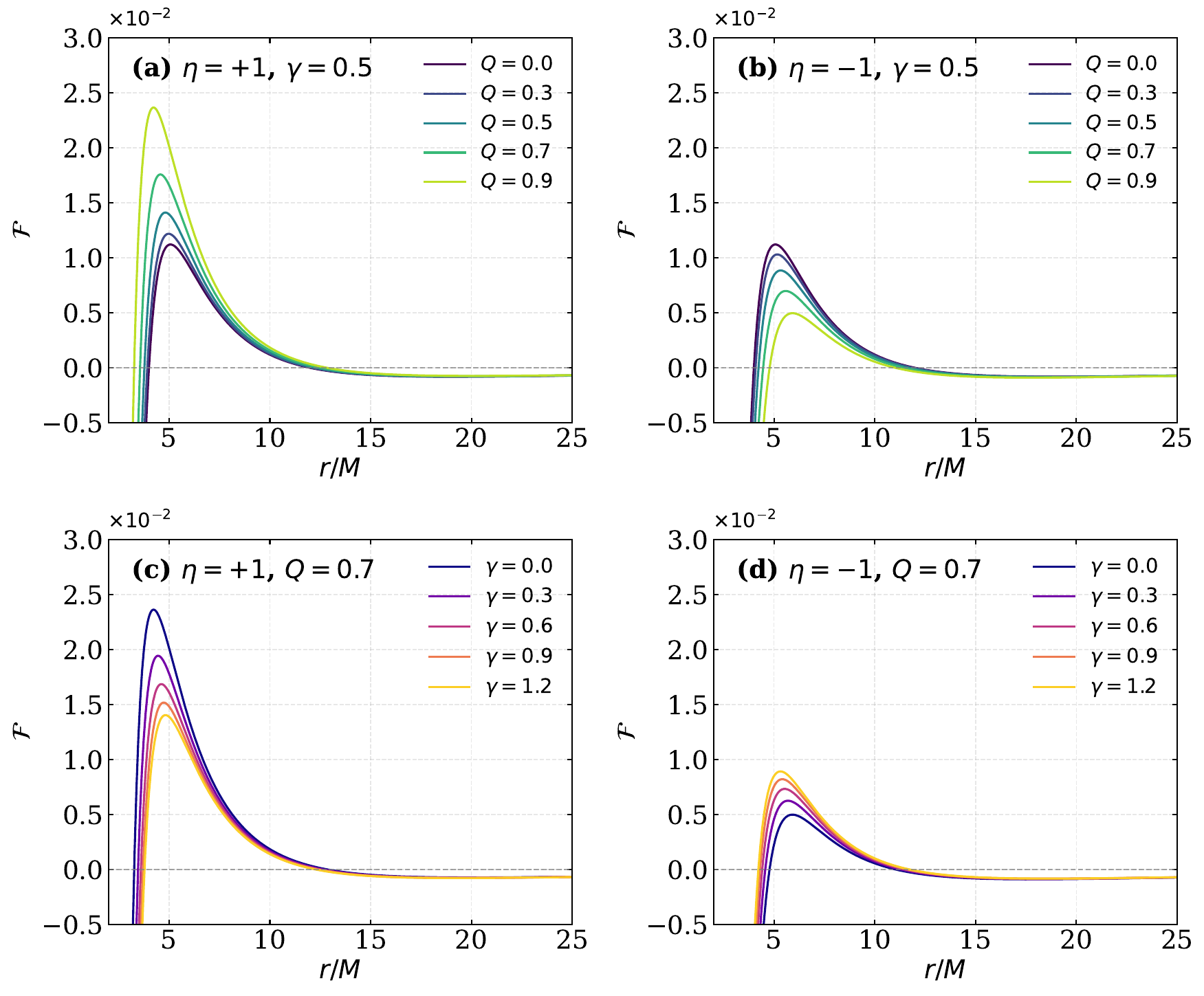}
    \caption{Effective force $\mathcal{F}$ acting on massive test particles as a function of the radial coordinate $r/M$ in the Mod(A)Max black hole spacetime. Figures (a) and (b) show the variation with electric charge $Q$ for fixed $\gamma = 0.5$, while Figs. (c) and (d) display the variation with the ModMax parameter $\gamma$ for fixed $Q = 0.7$. The left column corresponds to the ModMax case ($\eta = +1$) and the right column to the Mod(A)Max case ($\eta = -1$). The horizontal dashed gray line indicates $\mathcal{F} = 0$, where circular orbits occur. In all panels, $M = 1$ and $\mathcal{L} = 4$.}
    \label{fig:effective_force}
\end{figure}

In Fig.~\ref{fig:effective_force}, we present the effective radial force $\mathcal{F}$ acting on massive test particles in the equatorial plane of the Mod(A)Max black hole. The effective force is defined as the negative gradient of the effective potential, and its zeros correspond to the locations of circular orbits. When $\mathcal{F} > 0$, the net force is repulsive (outward), while $\mathcal{F} < 0$ indicates an attractive (inward) force. The horizontal dashed line at $\mathcal{F} = 0$ serves as a reference for identifying the radii of circular orbits.

Figures \ref{fig:effective_force}(a) and \ref{fig:effective_force}(b) illustrate the dependence of the effective force on the electric charge $Q$ for a fixed ModMax parameter $\gamma = 0.5$. In the ModMax case [Fig. \ref{fig:effective_force}(a), $\eta = +1$], increasing the charge $Q$ shifts the zero-crossing point (circular orbit radius) to smaller values of $r$, indicating that charged ModMax black holes allow stable circular orbits closer to the event horizon. The Mod(A)Max case [Fig. \ref{fig:effective_force}(b), $\eta = -1$] exhibits the opposite behavior: increasing $Q$ moves the zero-crossing to larger radii, suggesting that the phantom charge contribution effectively weakens the gravitational binding and pushes stable orbits outward.

Figures \ref{fig:effective_force}(c) and \ref{fig:effective_force}(d) display the influence of the ModMax coupling parameter $\gamma$ on the effective force for a fixed charge $Q = 0.7$. In the ModMax case [Fig. \ref{fig:effective_force}(c), $\eta = +1$], increasing $\gamma$ shifts the force curves upward, moving the circular orbit radius to larger values. For the Mod(A)Max case [Fig. \ref{fig:effective_force}(d), $\eta = -1$], larger values of $\gamma$ have the opposite effect, shifting the zero-crossing to smaller radii. These results are consistent with the behavior observed in the effective potential and confirm that the parameters $Q$ and $\gamma$ play competing roles in determining the orbital dynamics around Mod(A)Max black holes.

From the effective radial force, we observe that the geometric parameters, such as the black hole mass $M$, the electric charge $Q$, and the ModMax parameter $\gamma$, significantly influence its value. Additionally, the specific angular momentum $\mathcal{L}$ modifies this effective force.

Circular orbits on the equatorial plane satisfies the following conditions:
\begin{equation}
    U_{\rm eff}(r) = \mathcal{E}^2,\label{cc12}
\end{equation}
\begin{equation}
    \frac{\partial U_{\rm eff}}{\partial r} = 0.\label{cc13}
\end{equation}

Solving these equations yields the specific angular momentum and specific energy for test particles given by
\begin{equation}
    \mathcal{L}_{\rm sp} = \sqrt{\frac{r^3f'(r)}{2f(r)-rf'(r)}}=r\,\sqrt{\frac{\frac{M}{r}-\frac{\eta e^{-\gamma}\,Q^2}{r^2}}{1-\frac{3 M}{r}+\frac{2 \eta e^{-\gamma}\,Q^2}{r^2}}},\label{cc14}
\end{equation}
\begin{equation}
    \mathcal{E}_{\rm sp} =\sqrt{\frac{2f^2(r)}{2f(r)-rf'(r)}}=\pm\,\frac{\left(1-\frac{2 M}{r}+\frac{\eta e^{-\gamma}\,Q^2}{r^2}\right)}{\sqrt{1-\frac{3 M}{r}+\frac{2 \eta e^{-\gamma}\,Q^2}{r^2}}}.\label{cc15}
\end{equation}

We observe that the specific angular momentum and specific energy of test particles moving along circular orbits around a black hole depend on the black hole mass $M$, the electric charge $Q$, and the coupling parameter of the Mod(A)Max, $\eta\,e^{-\gamma}$. These quantities influence the spacetime curvature around the black hole and, consequently, significantly affect the motion and dynamics of the orbiting test particles.

\begin{itemize}
    \item When $\eta=+1$, ModMax black hole case, the specific angular momentum and specific energy simplify as
    \begin{equation}
        \mathcal{L}_{\rm sp} =r\,\sqrt{\frac{\frac{M}{r}-\frac{e^{-\gamma}\,Q^2}{r^2}}{1-\frac{3 M}{r}+\frac{2 e^{-\gamma}\,Q^2}{r^2}}},\qquad \mathcal{E}_{\rm sp} = \pm\,\frac{\left(1-\frac{2 M}{r}+\frac{e^{-\gamma}\,Q^2}{r^2}\right)}{\sqrt{1-\frac{3 M}{r}+\frac{2e^{-\gamma}\,Q^2}{r^2}}}.\label{case1}
    \end{equation}

    \item When $\eta=-1$, phantom ModMax or Mod(A) Max black hole case, the specific angular momentum and specific energy simplify as
    \begin{equation}
        \mathcal{L}_{\rm sp} =r\,\sqrt{\frac{\frac{M}{r}+\frac{e^{-\gamma}\,Q^2}{r^2}}{1-\frac{3 M}{r}-\frac{2 e^{-\gamma}\,Q^2}{r^2}}},\qquad \mathcal{E}_{\rm sp} = \pm\,\frac{\left(1-\frac{2 M}{r}-\frac{e^{-\gamma}\,Q^2}{r^2}\right)}{\sqrt{1-\frac{3 M}{r}-\frac{2e^{-\gamma}\,Q^2}{r^2}}}.\label{case2}
    \end{equation}
\end{itemize}

\begin{figure}[ht!]
    \centering
    \includegraphics[width=0.95\linewidth]{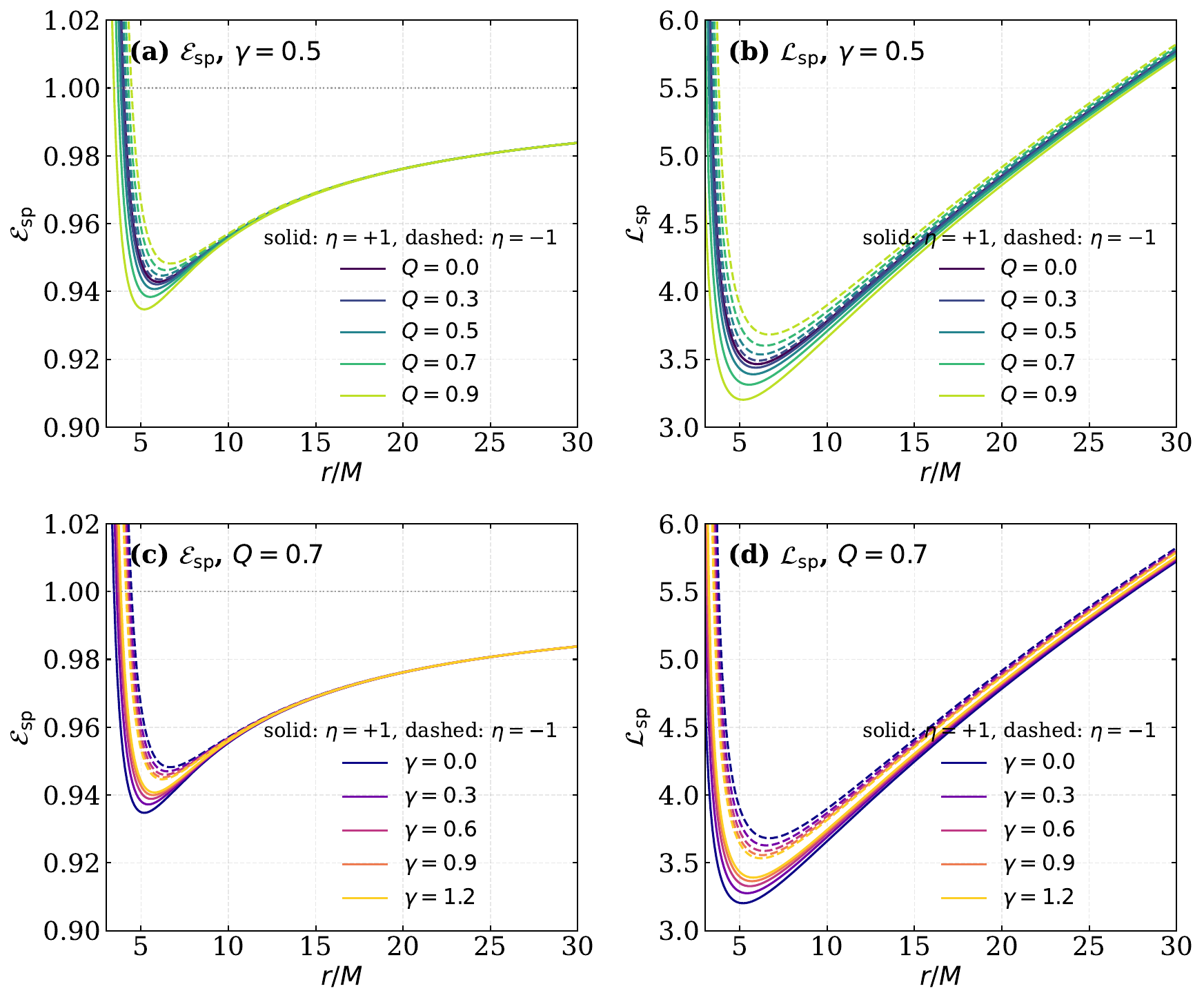}
    \caption{Specific energy $\mathcal{E}_{\rm sp}$ and specific angular momentum $\mathcal{L}_{\rm sp}$ for test particles in circular orbits as a function of the radial coordinate $r/M$ in the Mod(A)Max black hole spacetime. Figures (a) and (b) show the variation with electric charge $Q$ for fixed $\gamma = 0.5$, while Figs. (c) and (d) display the variation with the ModMax parameter $\gamma$ for fixed $Q = 0.7$. Solid lines correspond to the ModMax case ($\eta = +1$) and dashed lines to the Mod(A)Max case ($\eta = -1$). The horizontal dotted gray line in Figs. (a) and (c) indicates $\mathcal{E}_{\rm sp} = 1$, representing a particle at rest at infinity. Here $M = 1$.}
    \label{fig:E_L_specific}
\end{figure}

In Fig.~\ref{fig:E_L_specific}, we present the specific energy $\mathcal{E}_{\rm sp}$ and specific angular momentum $\mathcal{L}_{\rm sp}$ for test particles moving along circular orbits in the equatorial plane of the Mod(A)Max black hole. These quantities are fundamental for characterizing the orbital dynamics, as they determine the binding energy and rotational properties of particles at each orbital radius. The specific energy approaches unity as $r \to \infty$, corresponding to a marginally bound particle at rest at spatial infinity.

Figures \ref{fig:E_L_specific}(a) and \ref{fig:E_L_specific}(b) illustrate the dependence on the electric charge $Q$ for a fixed ModMax parameter $\gamma = 0.5$. For the specific energy [Fig. \ref{fig:E_L_specific}(a)], we observe that in the ModMax case ($\eta = +1$, solid lines), increasing the charge $Q$ leads to lower values of $\mathcal{E}_{\rm sp}$ at a given radius, indicating stronger gravitational binding. In contrast, the Mod(A)Max case ($\eta = -1$, dashed lines) shows the opposite trend: higher values of $Q$ result in larger $\mathcal{E}_{\rm sp}$, corresponding to weaker binding. The specific angular momentum [Fig. \ref{fig:E_L_specific}(b)] exhibits analogous behavior, with the ModMax case requiring smaller $\mathcal{L}_{\rm sp}$ for circular orbits at a given radius compared to the Mod(A)Max case.

Figures \ref{fig:E_L_specific}(c) and \ref{fig:E_L_specific}(d) display the influence of the ModMax coupling parameter $\gamma$ for a fixed charge $Q = 0.7$. In the ModMax case [solid lines], increasing $\gamma$ raises the specific energy and angular momentum curves, effectively reducing the binding strength. For the Mod(A)Max case [dashed lines], the effect is reversed: larger values of $\gamma$ lower the curves, enhancing the binding. These results demonstrate that the parameters $Q$ and $\gamma$ have opposite effects depending on the sign of $\eta$, providing a rich phenomenology for the orbital characteristics of test particles around Mod(A)Max black holes. The minimum of $\mathcal{L}_{\rm sp}$ corresponds to the innermost stable circular orbit (ISCO), whose location is significantly affected by the spacetime parameters.

The stability of circular orbits is determined by the second derivative of the effective potential. Stable orbits correspond to local minima ($\partial^2 U_{\rm eff}/\partial r^2 > 0$), while unstable orbits correspond to local maxima.  In Newtonian theory, the effective potential always has a minimum for any given angular momentum, and there are no ISCOs with a minimum
radius. However, when the form of the effective potential depends on the angular momentum of the particle and other parameters, the situation changes. The ISCO marks the boundary between stable and unstable circular orbits, defined by the conditions:
\begin{equation}
U_{\rm eff}(r) = \mathcal{E}^2, \qquad \frac{\partial U_{\rm eff}}{\partial r} = 0, \qquad \frac{\partial^2 U_{\rm eff}}{\partial r^2} \geq 0.\label{cc16}
\end{equation}

For marginally stable orbits, using Eq.~\eqref{cc10}, one can determine the ISCO radius using the following relation:
\begin{equation}
    \frac{3}{r}f(r)f'(r) - 2[f'(r)]^2 + f(r)f''(r) = 0.\label{cc17}
\end{equation}
Substituting the metric function $f(r)$ and after simplification, results the following cubic equation in $r$ as
\begin{equation}
    r^{3}
-6Mr^{2}
+9\,\eta e^{-\gamma}Q^{2} r
-4\,\eta^{2}e^{-2\gamma}Q^{4}/M
=0.\label{cc18}
\end{equation}

To solve the above cubic equation, we perform the transformation $r=x+2M$ and arrive at
\begin{equation}
    x^{3}+px+q=0,\label{cc19}
\end{equation}
where 
\begin{equation}
    p=9 \eta e^{-\gamma}Q^{2}-12M^{2}, \qquad
q=-16M^{3}+18 M \eta e^{-\gamma}Q^{2}-\frac{4 \eta^2 e^{-2\gamma} Q^{4}}{M}.\label{cc20}
\end{equation}
For this cubic equation, the discriminant is given by
\begin{equation}
 \Delta=\left(\frac{q}{2}\right)^{2}+\left(\frac{p}{3}\right)^{3}=\left(-8M^{3}+9M \eta e^{-\gamma}Q^{2}-\frac{2 \eta^2 e^{-2\gamma} Q^{4}}{M}\right)^{2}
+\left(3\eta e^{-\gamma}Q^{2}-4M^{2}\right)^{3}.\label{cc21}
\end{equation}

If $\Delta>0$, we have one real root for Eq. (\ref{cc19}) and the ISCO radius is given by
\begin{equation}
r_{\rm ISCO}=
2M
+\sqrt[3]{8M^{3}-9 M \eta e^{-\gamma} Q^2+\frac{2 \eta^2 e^{-2 \gamma} Q^2}{M}+\sqrt{\Delta}}
+\sqrt[3]{8M^{3}-9 M \eta e^{-\gamma}Q^{2}+\frac{2 \eta^2 e^{-2\gamma}Q^{4}}{M}-\sqrt{\Delta}}.\label{cc22} 
\end{equation}
On the other hand, if $\Delta=0$, all roots are real, with at least two equal.

We observe that the ISCO radius depends on the black hole mass $M$, the electric charge $Q$, and the coupling parameter of the Mod(A)Max, $\eta\,e^{-\gamma}$. In the limit $Q=0$, corresponding to the absence of the electric charge of the black hole, the space-time (\ref{metric}) to the standard Schwarzschild black hole, and the ISCO radius from Eq. (\ref{cc22}) simplifies to $r^{\rm Sch}_{\rm ISCO}=6 M$. 

\begin{figure}[ht!]
    \centering
    \includegraphics[width=0.95\linewidth]{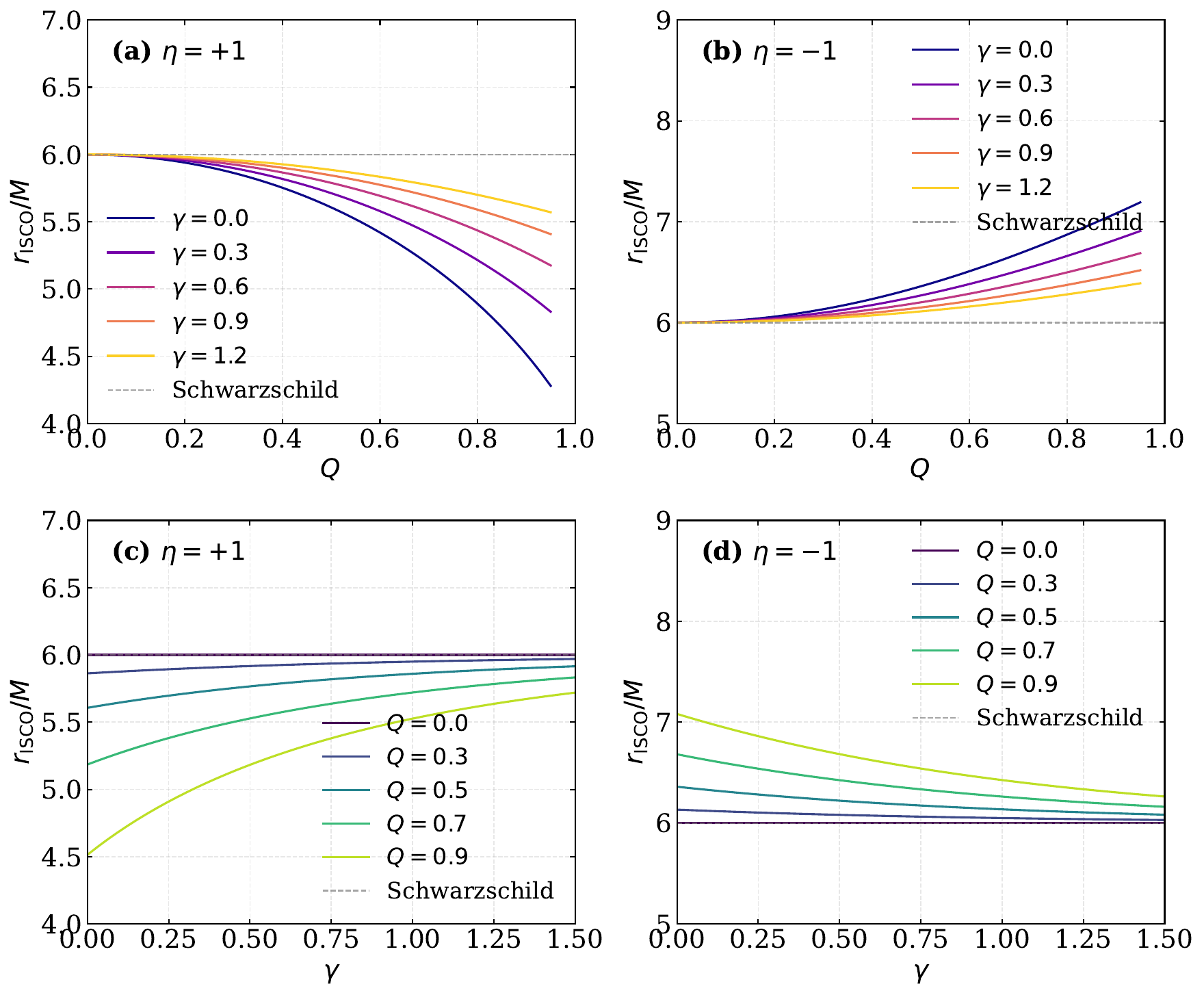}
    \caption{Innermost stable circular orbit (ISCO) radius $r_{\rm ISCO}$ as a function of the black hole parameters. Figures (a) and (b) show $r_{\rm ISCO}$ versus the electric charge $Q$ for different values of $\gamma$, while Figs. (c) and (d) display $r_{\rm ISCO}$ versus $\gamma$ for different values of $Q$. The left column corresponds to the ModMax case ($\eta = +1$) and the right column to the Mod(A)Max case ($\eta = -1$). The horizontal dashed gray line indicates the Schwarzschild value $r_{\rm ISCO} = 6M$. Here $M = 1$.}
    \label{fig:r_ISCO}
\end{figure}

In Fig.~\ref{fig:r_ISCO}, we present the innermost stable circular orbit (ISCO) radius as a function of the black hole parameters. The ISCO marks the boundary between stable and unstable circular orbits, and its location is of fundamental importance in accretion disk physics because it determines the inner edge of geometrically thin accretion disks and influences the efficiency of energy extraction from infalling matter.

Figures \ref{fig:r_ISCO}(a) and \ref{fig:r_ISCO}(b) show the dependence of $r_{\rm ISCO}$ on the electric charge $Q$ for various values of the ModMax parameter $\gamma$. In the ModMax case [Fig. \ref{fig:r_ISCO}(a), $\eta = +1$], increasing the charge $Q$ leads to a monotonic decrease in the ISCO radius, with the effect being more pronounced for smaller values of $\gamma$. This behavior indicates that the electric charge in the ModMax scenario allows stable orbits to exist closer to the black hole, potentially enhancing the radiative efficiency of accretion processes. In contrast, the Mod(A)Max case [Fig. \ref{fig:r_ISCO}(b), $\eta = -1$] exhibits the opposite trend: increasing $Q$ pushes the ISCO to larger radii, suggesting that the phantom charge contribution effectively weakens the gravitational binding and moves the innermost stable orbit outward.

Figures \ref{fig:r_ISCO}(c) and \ref{fig:r_ISCO}(d) illustrate the influence of the ModMax coupling parameter $\gamma$ on the ISCO radius for fixed values of $Q$. In the ModMax case [Fig. \ref{fig:r_ISCO}(c)], increasing $\gamma$ raises the ISCO radius toward the Schwarzschild value, indicating that larger $\gamma$ suppresses the charge-induced reduction of $r_{\rm ISCO}$. For the Mod(A)Max case [Fig. \ref{fig:r_ISCO}(d)], larger values of $\gamma$ lower the ISCO radius, partially compensating for the enhancement caused by the phantom charge. These results demonstrate that precise measurements of the ISCO location, for instance, through X-ray observations of accretion disks, could in principle provide constraints on the parameters of Mod(A)Max black holes.

\section{Quasiperiodic Oscillations (QPOs): Epicyclic Frequencies} \label{sec:5}

In this section, we investigate quasiperiodic oscillation (QPO) frequencies and analyze how the electric charge and the Mod(A)Max's parameter influence the azimuthal, radial, and vertical epicyclic frequencies of test particle motion. As discussed in the introduction, QPO frequencies play a crucial role in astrophysics because they are directly linked to the properties of compact objects and the structure of surrounding accretion disks. In particular, precise measurements of QPOs provide a powerful observational tool for probing the spacetime geometry near black holes, constraining their physical parameters, and testing deviations from general relativity. Understanding the dependence of QPO frequencies on additional parameters, such as electric charge and nonlinear electrodynamic effects encoded in the Mod(A)Max framework, therefore offers valuable insights into strong-field gravity and high-energy astrophysical processes.

To study the oscillatory motion of neutral particles, we perturb the equations of motion around stable circular orbits. When a test particle is slightly displaced from its equilibrium position is associated with a stable circular orbit in the equatorial plane, it exhibits epicyclic motion,
defined by linear harmonic oscillations.

It is well-known that the angular velocity measured by a distant observer is given by $\Omega_{\phi}= d\phi/dt$. For a spacetime of the form \eqref{metric}, this quantity can be written as
\begin{equation}
\Omega_{\phi}^2 = \frac{f'(r)}{2r}=\frac{M}{r^3}
-\frac{\eta e^{-\gamma}Q^2}{r^4}.
\label{ss1}
\end{equation}
where we have used relations (\ref{cc14})--(\ref{cc15}).

For fixed values of $(r,\theta)$, the angular frequency can be converted into a physical frequency measured in Hertz (Hz) according to 
\begin{equation}
\nu_{(\phi,r,\theta)} = \frac{1}{2\pi}\frac{c^3}{G M}\,\Omega_{(K,r,\theta)} \; \mathrm{[Hz]},
\label{ss2}
\end{equation}
which allows for direct comparison with astrophysical observations. In the above expression, we use $c = 3\times 10^8\,\mathrm{m\,s^{-1}}$ and $G = 6.67\times 10^{-11}\,\mathrm{m^3\,kg^{-1}\,s^{-2}}$.

\begin{center}
    {\bf II.\, Radial and Vertical Frequencies}
\end{center}

The radial and vertical angular frequencies \(\nu_r=\Omega_r\) and \(\nu_{\theta}=\Omega_{\theta}\) are the frequency of oscillations of the neutral test particles in the radial direction along the stable orbits, which can be determined from the second derivatives of the effective potential by $r$ and $\theta$ coordinates, respectively are \cite{ss27,ss28,ss29,ss30,ss31,ss32,ss32a,ss32b,ss32c,ss35,ss37,ss40,ss43,ss18,ss19,ss23}:
\begin{align} 
\Omega_r^2=-\frac{1}{2\, g_{rr}\,(u^t)^2}\, \frac{\partial^2 U_{\text {eff}}}{\partial r^2} \bigg{|}_{r =\mbox{const.}, \,\theta=\mbox{const.}}.\label{ss3} 
\end{align}
And
\begin{align} 
\Omega^2_{\theta}=- \frac{1}{2\, g_{\theta \theta }\, (u^t)^2} \frac{\partial ^2 U_{\text {eff}}}{\partial \theta ^2} \bigg{|}_{r =\mbox{const.}, \,\theta=\mbox{const.}}\ . \label{ss4}
\end{align}

Noted that these are the frequencies of neutral test particles as measured by a static distant observer. According to a local observer, the harmonic oscillatory motion frequencies are provided by \cite{ss35,ss37,ss40,ss43,ss18,ss19,ss23}: 
\begin{align}
    \omega_r^2&=-\frac{1}{2\, g_{rr}}\, \frac{\partial^2 U_{\text {eff}}}{\partial r^2} \bigg{|}_{r =\mbox{const.}, \,\theta=\mbox{const.}},\label{ss5} \\
    \omega^2_{\theta}&=-\frac{1}{2\, g_{\theta \theta }} \frac{\partial ^2 U_{\text {eff}}}{\partial \theta ^2} \bigg{|}_{r =\mbox{const.}, \,\theta=\mbox{const.}}\ . \label{ss6}
\end{align}
Here $\Omega^2_i=\frac{\omega^2_i}{(u^t)^2}$ with $u^t=\dot{t}=\sqrt{\frac{2}{2\,f(r)-r\,f'(r)}}$ is the relation between these two types of frequencies as measured by a local and a static distant observer. 

In our case at hand, the radial and vertical frequencies for a static distant observer are given by
\begin{align}
    \Omega^2_r&=-\frac{f(r)}{2}\left(\frac{3}{r}f(r)f'(r) - 2(f'(r))^2 + f(r)f''(r)\right)\nonumber\\
    &=-\frac{M}{r^3}
+\frac{8M^2}{r^4}
-\frac{15M\,\eta e^{-\gamma}Q^2}{r^5}
+\frac{10\,\eta^2 e^{-2\gamma}Q^4}{r^6}
-\frac{4M\,\eta^2 e^{-2\gamma}Q^4}{r^7}
+\frac{4\,\eta^3 e^{-3\gamma}Q^6}{r^8}.\label{ss7}
\end{align}
And
\begin{equation}
    \Omega^2_{\theta}=-\frac{f(r)\,f'(r)}{2\,r}=-\frac{M}{r^3}
+\frac{\eta e^{-\gamma}Q^2+2M^2}{r^4}
-\frac{3M\,\eta e^{-\gamma}Q^2}{r^5}
+\frac{\eta^2 e^{-2\gamma}Q^4}{r^6}.\label{ss8}
\end{equation}

From the epicyclic frequency expressions given in Eqs.~(\ref{ss1}), (\ref{ss7}), and (\ref{ss8}), we observe that the black hole mass $M$, the electric charge $Q$, and the Mod(A)Max's coupling term $\eta\,e^{-\gamma}$ significantly influence these frequencies when compared to those of a standard black hole. 
Furthermore, the parameters $\gamma$ and $\eta$ introduce additional deviations from the corresponding results for the Reissner-Nordström black hole, thereby modifying the azimuthal, radial, and vertical epicyclic frequencies. 
\begin{figure}[ht!]
    \centering
    \includegraphics[width=0.95\linewidth]{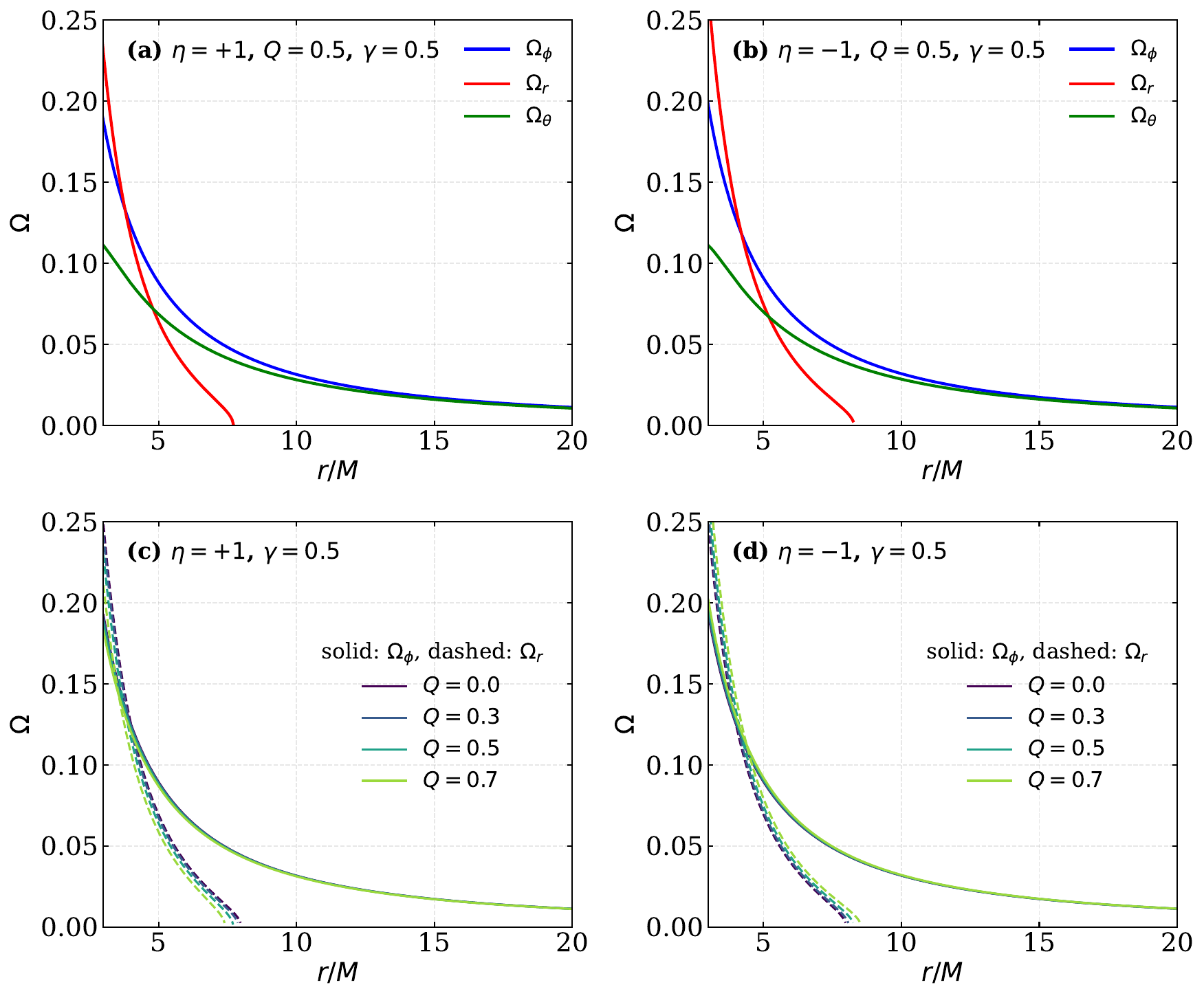}
    \caption{Epicyclic frequencies as a function of the radial coordinate $r/M$ in the Mod(A)Max black hole spacetime. Panels (a) and (b) show the azimuthal frequency $\Omega_{\phi}$ (blue), radial frequency $\Omega_{r}$ (red), and vertical frequency $\Omega_{\theta}$ (green) for fixed $Q = 0.5$ and $\gamma = 0.5$. Panels (c) and (d) display $\Omega_{\phi}$ (solid lines) and $\Omega_{r}$ (dashed lines) for different values of $Q$ with fixed $\gamma = 0.5$. The left column corresponds to the ModMax case ($\eta = +1$) and the right column to the Mod(A)Max case ($\eta = -1$). The radius where $\Omega_{r} = 0$ corresponds to the ISCO location. Here $M = 1$.}
    \label{fig:epicyclic_frequencies}
\end{figure}

In Fig.~\ref{fig:epicyclic_frequencies}, we present the epicyclic frequencies for test particles in circular orbits around the Mod(A)Max black hole. These frequencies characterize the oscillatory motion of particles perturbed from circular orbits: $\Omega_{\phi}$ is the azimuthal (Keplerian) frequency, $\Omega_{r}$ is the radial epicyclic frequency governing oscillations in the orbital plane, and $\Omega_{\theta}$ is the vertical epicyclic frequency describing oscillations perpendicular to the equatorial plane. These quantities are fundamental for understanding quasi-periodic oscillations (QPOs) observed in X-ray binaries and active galactic nuclei.

Figures \ref{fig:epicyclic_frequencies}(a) and \ref{fig:epicyclic_frequencies}(b) compare the three frequencies for the ModMax ($\eta = +1$) and Mod(A)Max ($\eta = -1$) cases with fixed $Q = 0.5$ and $\gamma = 0.5$. In both scenarios, the azimuthal frequency $\Omega_{\phi}$ dominates at all radii, while $\Omega_{r}$ vanishes at the ISCO radius, marking the transition from stable to unstable circular orbits. The vertical frequency $\Omega_{\theta}$ lies between these two. Notably, the frequencies in the Mod(A)Max case [Fig. \ref{fig:epicyclic_frequencies}(b)] are slightly lower than in the ModMax case [Fig. \ref{fig:epicyclic_frequencies}(a)] at the same radii, reflecting the different effective gravitational potentials.

Figures \ref{fig:epicyclic_frequencies}(c) and \ref{fig:epicyclic_frequencies}(d) illustrate the dependence of $\Omega_{\phi}$ and $\Omega_{r}$ on the electric charge $Q$ for fixed $\gamma = 0.5$. In the ModMax case [Fig. \ref{fig:epicyclic_frequencies}(c)], increasing $Q$ raises the frequencies at a given radius and shifts the zero of $\Omega_{r}$ (the ISCO) to smaller radii, consistent with the results shown in Fig.~\ref{fig:r_ISCO}. In contrast, for the Mod(A)Max case [Fig. \ref{fig:epicyclic_frequencies}(d)], increasing $Q$ lowers the frequencies and moves the ISCO outward. These distinct behaviors provide potential observational signatures that could distinguish between ModMax and Mod(A)Max black holes through the analysis of QPO frequencies in accreting systems.

\begin{center}
    {\bf II.\, Periastron Frequency}
\end{center}

To calculate the periastron precession, we consider a slight perturbation of the particle from its stable position, which leads to oscillations around that stable point characterized by a radial frequency $\Omega_r$. The periastron precession frequency ($\Omega_p$) is defined as the difference between the azimuthal (or orbital) frequency and the radial epicyclic frequency. Physically, this frequency characterizes the relativistic precession of the orbital ellipse of a test particle around the central compact object. Mathematically, it can be expressed as
\begin{equation}
    \Omega_p=\Omega_{\phi}-\Omega_r.\label{ss9}
\end{equation}
Unlike Newtonian gravity, where the radial and orbital frequencies are equal, general relativistic effects near a BH lead to the inequality $\Omega_{\theta} \neq \Omega_{\phi}$.

Substituting the azimuthal and radial frequencies from Eq. (\ref{ss1}) and (\ref{ss7}), we find the following expression:
\begin{equation}
\Omega_p=\sqrt{\frac{M}{r^3}
-\frac{\eta e^{-\gamma}Q^2}{r^4}}-\sqrt{-\frac{M}{r^3}
+\frac{8M^2}{r^4}
-\frac{15M\,\eta e^{-\gamma}Q^2}{r^5}
+\frac{10\,\eta^2 e^{-2\gamma}Q^4}{r^6}
-\frac{4M\,\eta^2 e^{-2\gamma}Q^4}{r^7}
+\frac{4\,\eta^3 e^{-3\gamma}Q^6}{r^8}}.\label{ss10}
\end{equation}

From the above expression, it is evident that the black hole mass $M$, the electric charge $Q$, and the Mod(A)Max's coupling term $\eta\,e^{-\gamma}$ alter this periastron frequency. 

\begin{figure}[ht!]
\centering
\includegraphics[width=0.95\linewidth]{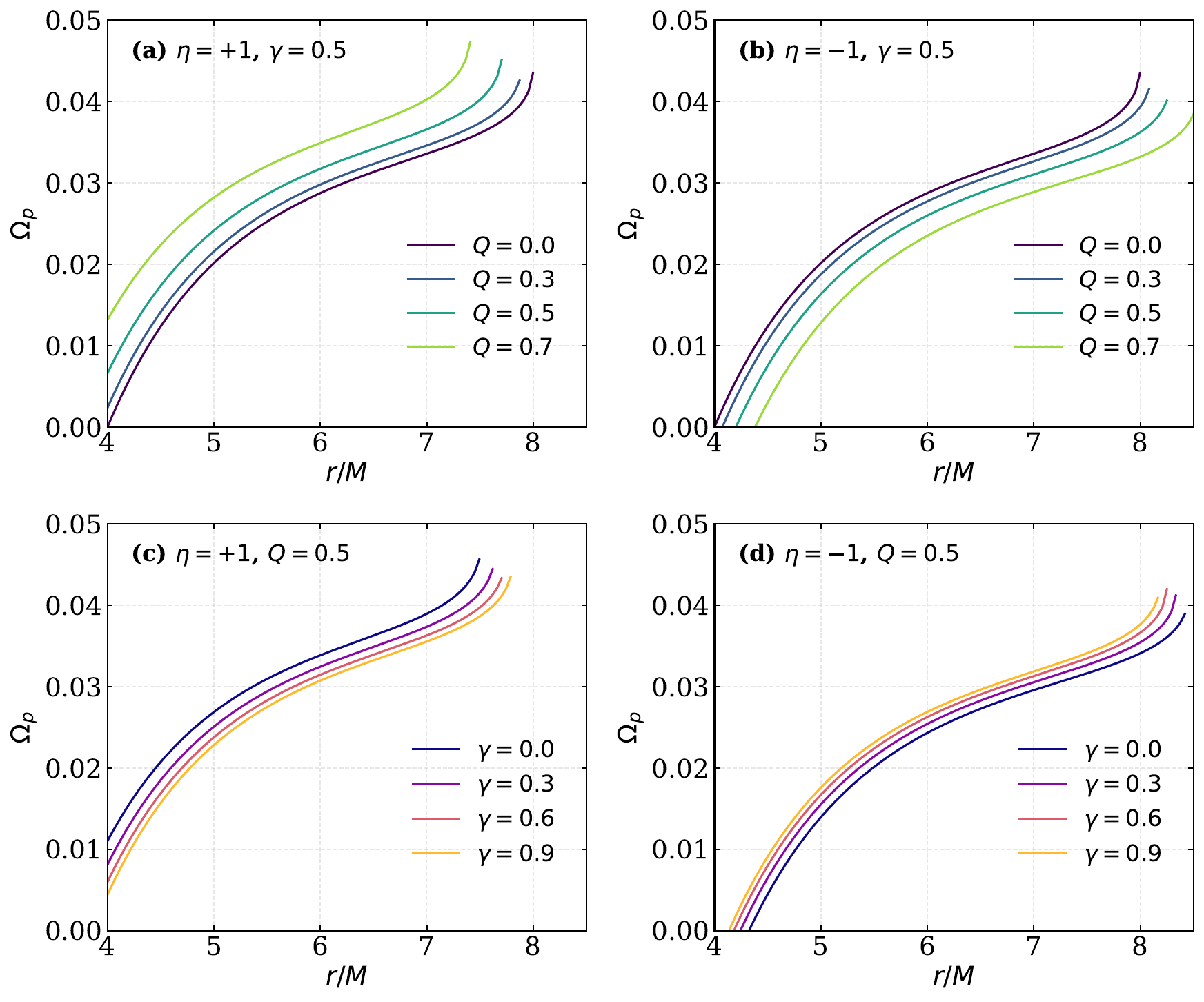}
\caption{Periastron precession frequency $\Omega_p$ as a function of the radial coordinate $r/M$ in the Mod(A)Max black hole spacetime. Panels (a) and (b) show the variation with electric charge $Q$ for fixed $\gamma = 0.5$, while panels (c) and (d) display the variation with the ModMax parameter $\gamma$ for fixed $Q = 0.5$. The left column corresponds to the ModMax case ($\eta = +1$) and the right column to the Mod(A)Max case ($\eta = -1$). Here $M = 1$.}
\label{fig:periastron_frequency}
\end{figure}
In Fig.~\ref{fig:periastron_frequency}, we present the periastron precession frequency (\ref{ss9}) for test particles in circular orbits around the Mod(A)Max black hole. The periastron precession frequency characterizes the relativistic precession of the orbital ellipse of a test particle around the central compact object. Unlike Newtonian gravity, where the radial and orbital frequencies are equal ($\Omega_r = \Omega_\phi$), general relativistic effects near a black hole lead to the inequality $\Omega_r \neq \Omega_\phi$, resulting in a net precession of the orbit.

Figures \ref{fig:periastron_frequency}(a) and \ref{fig:periastron_frequency}(b) illustrate the dependence of $\Omega_p$ on the electric charge $Q$ for a fixed ModMax parameter $\gamma = 0.5$. In the ModMax case [Fig. \ref{fig:periastron_frequency}(a), $\eta = +1$], increasing the charge $Q$ raises the precession frequency at a given radius, indicating that the electric charge enhances the relativistic precession effect. In contrast, the Mod(A)Max case [Fig. \ref{fig:periastron_frequency}(b), $\eta = -1$] exhibits the opposite behavior: increasing $Q$ lowers the precession frequency, suggesting that the phantom charge contribution partially counteracts the relativistic precession.

Figures \ref{fig:periastron_frequency}(c) and \ref{fig:periastron_frequency}(d) display the influence of the ModMax coupling parameter $\gamma$ on the periastron precession frequency for a fixed charge $Q = 0.5$. In the ModMax case [Fig. \ref{fig:periastron_frequency}(c), $\eta = +1$], increasing $\gamma$ lowers the precession frequency, counteracting the charge-induced enhancement observed in Fig. \ref{fig:periastron_frequency}(a). For the Mod(A)Max case [Fig. \ref{fig:periastron_frequency}(d), $\eta = -1$], larger values of $\gamma$ raise the precession frequency, partially compensating the suppression caused by the phantom charge. These distinct behaviors provide potential observational signatures that could be used to constrain the parameters of Mod(A)Max black holes through precise measurements of orbital precession in systems such as the S-stars orbiting Sgr A* or in X-ray binary pulsars.

\section{Conclusion}\label{sec:6}

In this work, we investigated the observable signatures, particle dynamics properties, and QPOs around a Mod(A)Max black hole. The spacetime geometry, described by the metric function in Eq.~\eqref{function}, incorporates four parameters: the BH mass $M$, the electric charge $Q$, the Mod(A)Max's parameter $\gamma$, and the coupling parameter $\eta$. This coupling between a new modification of Maxwell theory (ModMax) and the anti-Maxwell (phantom) field within the Einstein–Hilbert action allowed us to explore the classical black hole solutions and their observable signatures. 

In Sec.~\ref{sec:3}, we studied null geodesics and the photon sphere radius properties using the Lagrangian formalism. First, we obtain the effective potential for photons, given in Eq.~\eqref{bb7}. In Fig.~\ref{fig:Veff_photon}, we show that at $\eta=1$, both $Q$ and $\gamma$ suppress the potential peak, lowering the barrier height and allowing photons with lower impact parameters to escape. However, in the situation of phantom $\eta=-1$, both $Q$ and $\gamma$ enhance the potential peak, hence increasing the barrier height. The photon sphere radius $r_s$ given in Eq.~\eqref{bb11}, decreases monotonically with both $Q$ and $\gamma$ in the case of $\eta=1$, as demonstrated in Fig.~\ref{fig:photon-radius}. In contrast, the Mod(A)Max case $\eta=-1$ exhibits the opposite trend. Regarding the BH shadow observable by distant observers, we used Eq.~\eqref{bb14}, which is acceptable for asymptotically flat spacetimes. The shadow radius $R_{sh}$ decreases with both parameters (both $Q$ and $\gamma$ (Fig.~\ref{fig:shadow-radius}), in the case of ordinary ModMax, whereas it grows in the case of Mod(A)Max. Overall, depending on the sign of $\eta$, the shadow radius might be bigger (in the case of $\eta=-1$) or less (in the case of $\eta=1$) than the Schwarzschild case. We also determined the photon trajectories and analyzed how various parameters alter their trajectory.  The interplay between the electric charge $Q$ and the parameter $\eta$ plays a crucial role in shaping photon dynamics around the black hole. The critical impact parameter, which separates capture from scattering, directly determines the observed shadow radius. Consequently, the ModMax and Mod(A)Max frameworks may produce distinguishable imprints in black hole shadows and gravitational lensing observations (see Fig. \ref{fig:photon_trajectory}).

In Sec .~\ref {sec:4}, we investigated the dynamics of neutral test particles around a Mod(A)Max black hole. In Fig. \ref{fig:Ueff_massive}, we showed the effective potential for a massive test particle and analyzed how the spacetime parameters influence the potential. Next, we determined the effective force experienced by the test particle. Figure \ref{fig:effective_force} displayed that for fixed charge $Q$, the ModMax coupling parameter$\gamma$ significantly alters the effective force and hence the circular photon orbits. In the ModMax case ($\eta=1$), increasing $\gamma$ shifts the orbit radius outward, while in the Mod(A)Max scenario ($\eta=-1$) it shifts inward. This contrasting behavior confirms the competing roles of $Q$ and $\gamma$ in governing orbital dynamics around Mod(A)Max black holes. Figure \ref{fig:E_L_specific} illustrates the dependence of specific $\mathcal{E}_{\rm sp}$ and specific angular momentum $\mathcal{L}_{\rm sp}$   for test particles in circular orbits as a function of the radial coordinate.  For a fixed charge $Q$, the ModMax coupling $\gamma$ has opposite effects on particle energetics depending on the sign of $\eta$. In the ModMax case, increasing  $\gamma$ raises the specific energy and angular momentum, weakening orbital binding, whereas in the Mod(A)Max scenario, it lowers them, enhancing stability. Consequently, the ISCO location identified by the minimum of $\mathcal{L}_{\rm sp}$ is strongly sensitive to the spacetime parameters.

In Sec.~\ref{sec:5}, we investigated QPOs, which serve as probes of strong gravity near compact objects. We calculated the azimuthal frequency $\Omega_\phi$  and found that it relies on $Q$, $\gamma$, and $\eta$. In Fig. \ref{fig:epicyclic_frequencies}, epicyclic frequencies reveal distinct orbital dynamics in ModMax and Mod(A)Max black holes, with the azimuthal frequency dominating and the radial frequency vanishing at the ISCO. While increasing the charge $Q$ enhances the frequencies and shifts the ISCO inward in the ModMax case, it suppresses them and moves the ISCO outward in the Mod(A)Max scenario. These contrasting behaviors may provide observable signatures in QPO spectra, allowing discrimination between the two models.

\footnotesize

\section*{Acknowledgments}

F.A. acknowledges the Inter University Centre for Astronomy and Astrophysics (IUCAA), Pune, India, for granting a visiting associateship. E. O. Silva acknowledges the support from Conselho Nacional de Desenvolvimento Cient\'{i}fico e Tecnol\'{o}gico (CNPq) (grants 306308/2022-3), Funda\c c\~ao de Amparo \`{a} Pesquisa e ao Desenvolvimento Cient\'{i}fico e Tecnol\'{o}gico do Maranh\~ao (FAPEMA) (grants UNIVERSAL-06395/22), and Coordena\c c\~ao de Aperfei\c coamento de Pessoal de N\'{i}vel Superior (CAPES) - Brazil (Code 001).

\section*{Data Availability Statement}

This manuscript has no associated data [Author's comment: No data were generated or analyzed in this manuscript].

\section*{Code/Software}

No new code/software were developed in this manuscript [Authors comment: No new code/software were developed in this manuscript].

\end{document}